%% file: BPH-24-003_temp.tex
\pdfoutput=1
\documentclass[11pt,twoside,a4paper,cmspaper,final,collab]{cms-tdr}
\def\svnVersion{b54a32c-D}\def\svnDate{2026/02/01}\def\cmsCernNoTag{CERN-EP-2025-248}\def\cmsCernDate{\today}\def\cmsMessage{Submitted to Science Advances}
\begin{document}\cmsNoteHeader{BPH-24-003}

\newcommand{\JJ}{\ensuremath{\PJGy\,\PJGy}\xspace}
\newcommand{\Jpsi}{\PJGy}
\newcommand{\JPp}{\ensuremath{\PJGy\,\PGyP{2S}}\xspace}
\newcommand{\Pps}{\PGyP{2S}}
\newcommand{\Xss}{\PXP{6600}}
\newcommand{\Xsn}{\PXP{6900}}
\newcommand{\Xso}{\PXP{7100}}
\newcommand{\nbin}{\ensuremath{n_\text{bin}}\xspace}

\newcommand{\PADstz}{{\HepAntiParticle{\PD}{}{\ast0}}\xspace}
\newcommand{\PZPp}[1]{{\HepParticleResonanceFull{\PZ}{}{}{#1}{}{+}}\xspace}
\newcommand{\PAGXcc}{{\HepAntiParticle{\PGX}{cc}{}}\xspace}
\newcommand{\PAGXmcc}{{\HepParticle{\PGX}{\PQc\PQc}{-}}\xspace}
\newcommand{\PAGXmmcc}{{\HepParticle{\PGX}{\PQc\PQc}{--}}\xspace}

\ifthenelse{\boolean{cms@external}}{\newcommand{\insupp}[2]{#1}}{\newcommand{\insupp}[2]{#2}}

\hyphenation{pa-ra-me-ter-i-za-tion}
\hyphenation{tetra-quark}
\hyphenation{tetra-charm}

\cmsNoteHeader{BPH-24-003}

\title{Observation of a family of all-charm tetraquarks}

\date{\today}

\abstract{
Three structures, \Xss, \Xsn, and \Xso, have emerged from the \JJ ($\Jpsi \to \PGmp\PGmm$) mass spectrum. These are candidates of all-charm tetraquarks, an exotic form of hadronic matter. A clearer picture of these states is obtained using proton-proton collision data collected by the CMS detector that corresponds to 315\fbinv, which yields 3.6 times more \JJ pairs than previous studies by CMS. All three structures, and their mutual interference, have statistical significances above five standard deviations. The presence of interference implies that the structures have common quantum numbers. Their squared masses align linearly with a resonance index and have natural widths that systematically decrease as the index increases. These features are consistent with radial excitations of tetraquarks composed of two aligned spin-1 diquarks without orbital excitation, and disfavor other interpretations. The $\JPp \to \PGmp\PGmm\PGmp\PGmm$ decay mode is also explored and the \Xsn and \Xso states are found with significances exceeding 8 and 4 standard deviations, respectively.
}

\author{{CMS Collaboration}}
\hypersetup{
pdfauthor={CMS Collaboration},
pdftitle={Observation of a family of all-charm tetraquarks},
pdfsubject={CMS},
pdfkeywords={CMS, tetraquark, exotic hadron, charmonium}} 

\maketitle

\section{\scihead{Introduction}}
The proliferation of ``elementary'' particles in the 1950s prompted Gell-Mann~\cite{GellMann} and Zweig~\cite{Zweig} to independently propose in 1964 that strongly-interacting particles (hadrons) are composed of either quark/antiquark pairs ($\PQq\PAQq$ mesons) or quark triplets ($\PQq\PQq\PQq$ baryons). 
Initially viewed as mathematical constructs, quarks gained general acceptance as physical entities, especially because of the 1974 discovery of the charm (\PQc) quark in the \Jpsi ($\PQc\PAQc$) meson (with a mass near 3100\MeV)~\cite{JPsiMIT,JPsiSLAC}, and the subsequent discoveries of other hadrons containing charm quarks.
Gell-Mann and Zweig also observed that symmetry principles permitted tetraquark $(\PQq\PAQq\PQq\PAQq)$ and pentaquark $(\PQq\PQq\PQq\PQq\PAQq)$ states. 
Although the existence of exotic hadrons, defined as systems other than $\PQq\PAQq$ mesons and $\PQq\PQq\PQq$ baryons, has long been considered, compelling evidence remained elusive. 
In the light-quark (\PQu, \PQd, \PQs) sector, broad resonance widths and overlap between exotic and conventional states prevented clear identification.
By the 1990s, exotics were plausible, but lacked unambiguous confirmation~\cite{Landsberg:1999Exotic}.

The search for exotics composed of heavy-quarks (\PQc, \PQb) has two advantages compared to the light-quark sector.  First, with quark masses significantly larger than the characteristic energy scale of the strong interaction $(\Lambda_\mathrm{QCD} \approx 200\MeV)$, more reliable theoretical treatments are available.  Second, the mass differences (or splittings) between different states are expected to be larger than the widths, allowing individual states to be more easily resolved.
The turning point occurred in 2003 with the discovery of the $\PXP{3872}$ resonance in the $\Jpsi\,\PGpp\PGpm$ decay channel~\cite{Belle:2003nnu}. 
Although resembling an excited charmonium ($\ccbar$) state, its mass was inconsistent with expectations. 
Lying near the $\PDz \PADstz$ threshold, it has been argued that $\PXP{3872}$ is a loosely bound $\PDz\PADstz$ ([{\PQc}{\PAQu}][{\PQu}{\PAQc}]) molecular state~\cite{Close:2003X3872}. 
Numerous heavy exotic hadrons followed~\cite{Ali:2019roi,Chen:2022ExoticRev}, including the charged charmonium-like $\PZPp{3930} ([{\PQc}{\PAQd}][{\PQu}{\PAQc}])$ decaying to $\Jpsi\,\PGpp$~\cite{BESIII-Z,Belle-Z} and the pentaquark state $\mathrm{P}_{\ccbar}(4312)^+$ decaying to $\Jpsi\,\Pp$~\cite{LHCb:2015yax}. 
 
The key to establishing the existence of exotic hadrons was the presence of a heavy-quark component, and systems composed entirely of heavy quarks promise even greater clarity.
With ever-increasing data samples, the first all-heavy tetraquark candidates have now been reported. 
Growing evidence has emerged for three tetracharm ($\mathrm{T}_{\PQc\PQc\PAQc\PAQc}$) candidates, \Xss, \Xsn, and \Xso, all decaying to \JJ ($\Jpsi \to \Pgmp\Pgmm$).
The \Xsn was the first structure found, by the LHCb Collaboration in 2020~\cite{LHCb:2020bwg}, and has also been observed by the ATLAS~\cite{ATLAS:2023bft} and CMS~\cite{cmscollaboration2023observation} Collaborations, all with statistical significances above five standard deviations $(5\sigma)$.  The \Xss was discovered in 2024 by the CMS Collaboration with a significance approaching $8\sigma$~\cite{cmscollaboration2023observation}, although a similar structure appears in the data previously published by LHCb~\cite{LHCb:2020bwg} and ATLAS~\cite{ATLAS:2023bft}.
Evidence for the \Xso is less conclusive as CMS reported a significance of $4.7\sigma$, while LHCb noted a small excess around 7200\MeV, but did not consider it further because of its low significance~\cite{LHCb:2020bwg} and ATLAS has seen no evidence~\cite{ATLAS:2023bft}. ATLAS has also reported two searches for tetracharm states in the \JPp decay mode~\cite{ATLAS:2023bft,ATLAS:J2S-2025}, with the \Xsn clearly seen at a significance well above $5\sigma$, while the \Xso state appeared at the $3\sigma$ level in one model in the first search using just $\PGyP{2S}\to\PGmp\PGmm$ decays~\cite{ATLAS:2023bft}, but was not found in the second search, which added $\PGyP{2S}\to\Jpsi \PGpp \PGpm (\Jpsi \to \PGmp\PGmm)$ decays~\cite{ATLAS:J2S-2025}.

There are still unanswered questions.
The existence of the \Xso hadron needs to be firmly established.
The ATLAS, CMS, and LHCb Collaborations all reported that fits including interference effects provided a better description of the data, but the specific fit models of the mass spectrum used were not mutually consistent, and the presence and nature of the interference was not conclusive.
In addition, all experiments reported an unexplained excess of \Jpsi pairs just above the \JJ threshold~\cite{cmscollaboration2023observation,LHCb:2020bwg,ATLAS:2023bft}.

\begin{figure*}[htb]
    \centering
    \includegraphics[width=0.95\textwidth]{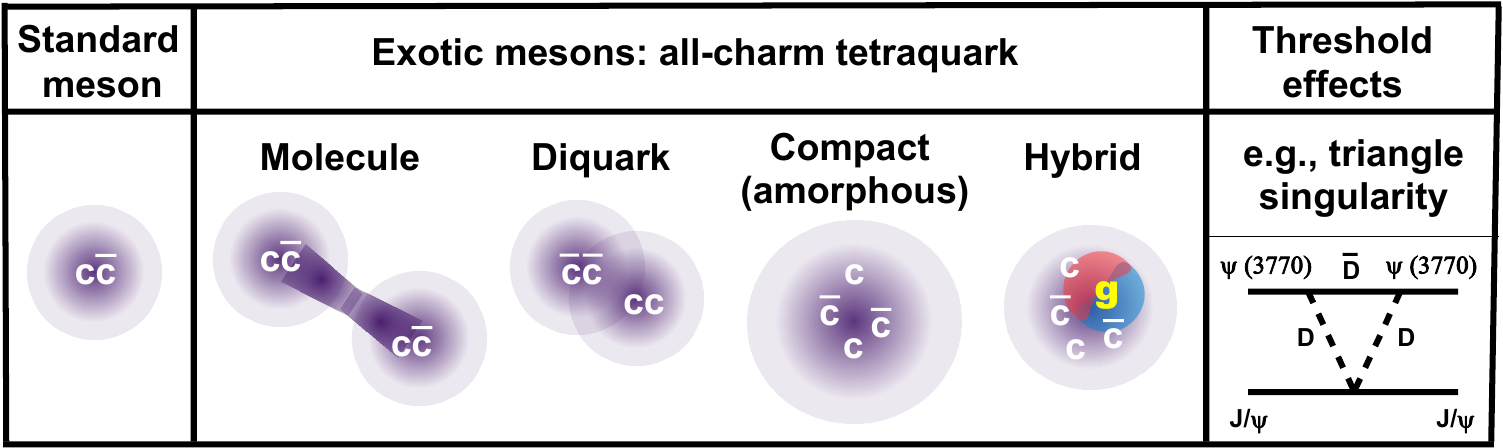}
    \caption{\firstsentence{Idealized models of all-charm structures.}
    Far left: conventional charmonium state.
    Center: various tetraquark configurations, \ie, [$\PQc\PAQc$][$\PQc\PAQc$] molecule, [$\PQc\PQc$][$\PAQc\PAQc$] diquarks, compact tetraquark with amorphous substructure, and hybrid $\PQc\PQc\PAQc\PAQc\Pg$ state.
    Far right: example of a nonresonant threshold effect, a ``triangle singularity'', where virtual scattering of \Jpsi and \PGyP{3770} mesons is enhanced through a triangular loop exchanging $\PD$ mesons, potentially leading to a peak-like structure in the \JJ mass spectrum around 6900\MeV~\cite{Lu:2023Triangle}.
    }
    \label{HadronCartoon}
\end{figure*}

The paramount question is the following.
What is the nature of these structures in the \JJ mass spectrum? 
Are they tetraquarks and, if so, of what sort?
In principle, there is a continuum of possible internal configurations of four-quark systems, but attention is usually focused on a few idealized types, illustrated in Fig.~\ref{HadronCartoon}.
The commonly considered models are:
a loosely-bound meson-meson ([$\PQc\PAQc$][$\PQc\PAQc$]) ``molecule''~\cite{Close:2003X3872,Santowsky:2021bhy}, a pair of ``diquarks'' ([$\PQc\PQc$][$\PAQc\PAQc$])~\cite{Debastiani:DiquarkSize2017,PhysRevD.102.114039,Barabanov:2020diquarks,Zhu:2020xni,Wang:2021kfv,PhysRevD.71.014028,Wang:2017jtz}, an amorphous system with no substructure~\cite{Chen:2022ExoticRev,Gordillo:2020sgc}, or a hybrid where there is also a valence gluon (an active, constituent gluon)~\cite{Wan:2020fsk,tang2025}. 
Diquarks are $\PQq\PQq$ or $\PAQq\PAQq$ pairs that are thought to form a quasi-particle because of a small quark-quark attraction.
Diquarks have been used in baryon models, but direct evidence of their existence remains inconclusive~\cite{Barabanov:2020diquarks}. 
In parallel, some interpretations attribute the observed structures not to genuine resonances but to nonresonant dynamic effects near dicharmonium thresholds~\cite{Guo:2019twa,Gong:2020bmg,Dong:2020,Bugg:2008Cusp,Swanson:2014CouplChann,Szczepaniak:2015eza,Chen:2022ExoticRev,Lu:2023Triangle}, an example of which is a triangle singularity~\cite{Lu:2023Triangle}, depicted in Fig.~\ref{HadronCartoon}.

In this article, we extend our previous study~\cite{cmscollaboration2023observation} with data collected in 2022--2024, which yield a sample with 3.6 times more \JJ pairs than previously available.
We also perform a study of $\JPp \to \PGmp\PGmm\PGmp\PGmm$ events.
The \Pps is an excited state of the \Jpsi, sharing the same quantum numbers, making \JPp a likely \PX decay mode.   The results of this analysis are also tabulated in the HEPData record~\cite{hepdata}.

\section{\scihead{Results}}
\subsection{Data selection} 
Our previous analysis~\cite{cmscollaboration2023observation} used data from proton-proton ($\Pp\Pp$) collisions at a center-of-mass energy ($\sqrt{s}$) of 13\TeV from Run~2 (2016--2018) at the CERN Large Hadron Collider (LHC), corresponding to an integrated luminosity (a measure of the total number of proton-proton collisions) of 135\fbinv. 
In this work, we add parking data~\cite{CMS:2024zhe} from Run~3 (2022--2024) at $\sqrt{s} = 13.6\TeV$, corresponding to an additional integrated luminosity of 180\fbinv.  Details of the data selection are provided in the Materials and Methods section.

\begin{figure*}[htb]
\centering
\includegraphics[width=0.95\textwidth]{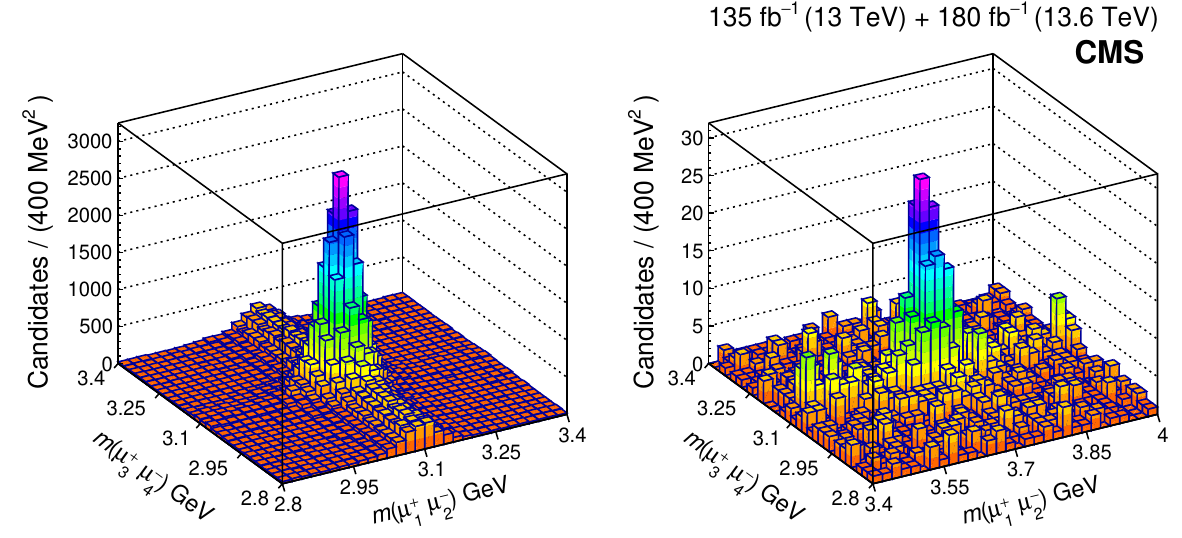}
\caption
{
    \firstsentence{The 2D event distributions in the $m(\PGm_1^{+},\PGm_2^{-})$ \vs $m(\PGm_3^{+},\PGm_4^{-})$ plane.} 
    The two opposite-sign muon pairs, $(\PGm_1^{+}\PGm_2^{-})$ and $(\PGm_3^{+}\PGm_4^{-})$, are ordered by the dimuon transverse momenta in \JJ events, and by the dimuon mass in \JPp events.
    The \JJ (left) and \JPp (right) events with four-muon invariant masses below 15\GeV for Run~2+3 data are shown. 
}
\label{3Dscattering}
\end{figure*}

The selection of \JJ events in this analysis follows the same criteria as used in the previous study~\cite{cmscollaboration2023observation}, applied to both the Run~2 and Run~3 data sets.
A two-dimensional (2D) distribution of the two dimuon masses for the selected \JJ candidates, limited to a double-\Jpsi mass below 15\GeV, is shown in Fig.~\ref{3Dscattering} (left). 
A 2D unbinned maximum likelihood fit is performed, where the \Jpsi signal on each axis is modeled with the sum of two Crystal Ball functions~\cite{CrystalBallRef}, and the nonresonant dimuon background is described by an independent first-order Chebyshev polynomial along each axis.
In the 6--15\GeV \JJ mass range, fits to the Run~2 and Run~3 data return signal yields of $12\,622 \pm 165$ and $31\,802 \pm 476$, respectively.
The total yield is about 3.6 times larger than that of Run~2, benefiting from a moderate (33\%) increase in integrated luminosity and a substantial improvement in trigger efficiency, albeit with a larger fraction of background events.

As the $\PGyP{2S}$ dimuon branching fraction is much smaller than for \Jpsi, even though their topologies are very similar, the \JJ selection criteria are reoptimized for the $\JPp \to \PGmp\PGmm\PGmp\PGmm$ events.
This reoptimization is achieved by adjusting requirements on the minimum transverse momentum (\pt) of the \Jpsi and \Pps candidates, the minimum \pt and identification requirements of the muons from the \Pps decay, and the invariant mass requirements for both \Jpsi and \Pps.
Figure~\ref{3Dscattering} (right) shows the 2D distribution of \Pps versus \Jpsi masses in \JPp events.
A 2D unbinned maximum likelihood fit to the Run~2+3 data using the same functional forms as for the \JJ fit yields $386 \pm 26$ \JPp signal events with \JPp invariant mass spectrum up to 15\GeV.

\subsubsection{The \texorpdfstring{\JJ}{J/psi J/psi} and \texorpdfstring{\JPp}{J/psi psi(2S)} mass spectra and signal structures}
The 1D invariant mass distributions for \JJ and \JPp events are fit by maximizing the unbinned extended-likelihood function~\cite{BARLOW1990496} from their respective thresholds of 6194\MeV for \JJ and 6783\MeV for \JPp up to 15\GeV.
The resonance signals are modeled by relativistic S-wave (\ie, of zero orbital angular momentum, $L$) Breit--Wigner (BW) functions (Section 50.3.1 of Ref.~\cite{PhysRevD.110.030001}).
The relativistic BW function includes a phase-space factor that reflects the kinematic turn-on above the resonance threshold, which depends on the masses of the parent and daughter particles.

The background model for the \JJ events includes five sources.
The first component, the near-threshold enhancement reported by all LHC experiments~\cite{cmscollaboration2023observation,LHCb:2020bwg,ATLAS:2023bft}, is modeled by a BW function ($\mathrm{BW}_0$) and incorporated as a background component.
The second contribution originates from nonresonant single-parton scattering (NRSPS) in which two \Jpsi mesons are produced from a single parton-parton interaction.  
This contribution is modeled using Monte Carlo (MC) simulated events.  
The third background component is double-parton scattering (DPS), which arises from two independent parton interactions within a single pp collision, producing uncorrelated \Jpsi meson pairs.  
This component is modeled using event mixing in which \Jpsi mesons reconstructed in different data events are combined.  
The fourth background source is feed-down contributions from $\PX \to \JPp$ decays, where one \Jpsi originates from the decay of \Pps, \eg, $\PGyP{2S} \to \PJGy\,\PGpp\PGpm$.  
Feed-down contributions from both \Xsn and \Xso decays are simulated.  
The fifth background arises from muon candidate combinations not originating from a \Jpsi meson.
The four-muon mass shape of this background is estimated from the dimuon invariant mass sidebands using the ``nine-tile'' method~\cite{CMD-3:2019cvx}, while the yield is obtained from the 2D fit to the $m(\Jpsi_1)$ versus $m(\Jpsi_2)$ distribution and included in the final fit as a Gaussian constraint. 
Each non-BW background component is parameterized separately by the product of an exponential, a polynomial, and a power-law term, using the four-muon invariant mass minus the threshold mass as the independent parameter. In the final fit of the di-\Jpsi mass spectrum, all of the background shape parameters are fixed to the values found in the estimation just described except for the exponential slope parameter of the NRSPS contribution, which is left free.

To fit the \JJ invariant mass spectrum, we extend the Run~2 baseline model~\cite{cmscollaboration2023observation}, which consists of the three $\PX$ signals plus the NRSPS, DPS, and threshold background terms, by explicitly adding the combinatorial background and the feed-down contribution.
The three resonant structures are described using the ``three-way'' BW interference model already employed in the Run~2 analysis~\cite{cmscollaboration2023observation}.
Interference is introduced through a term proportional to $\abs{\mathcal{M}}^{2}=\abs{r_1 \, e^{i\phi_1}\mathrm{BW}_{6600} + \mathrm{BW}_{6900} + r_3 e^{i\phi_3}\mathrm{BW}_{7100}}^2$, where $r_j$ and $\phi_j$ are the relative amplitudes and phase angles, as only phase angle differences are physically relevant.
Detector resolution and efficiency effects are excluded from the default \JJ model, as signal MC studies show that their impacts are small for masses well above threshold; the relative mass resolution varies from 0.15\% at 6500\MeV to 0.30\% at 7300\MeV.
Both effects are evaluated and incorporated as systematic uncertainties.
The near-threshold background, $\mathrm{BW}_0$, is treated as a noninterfering BW function, and is implemented with a term proportional to the square of its amplitude convolved with the Gaussian resolution function, $\abs{\mathrm{BW}_0}^2\otimes R(M_0)$, where $R(M_0)$ is evaluated at the central value $M_0$ of $\mathrm{BW}_0$ and $\otimes$ indicates convolution. 
As documented in the Supplementary Material, we also consider a noninterference model where none of the fit components interfere, \ie, each BW term is proportional to the square of its own amplitude convolved with the resolution function.

\begin{figure}[htb]
\centering
\includegraphics[width=0.49\textwidth]{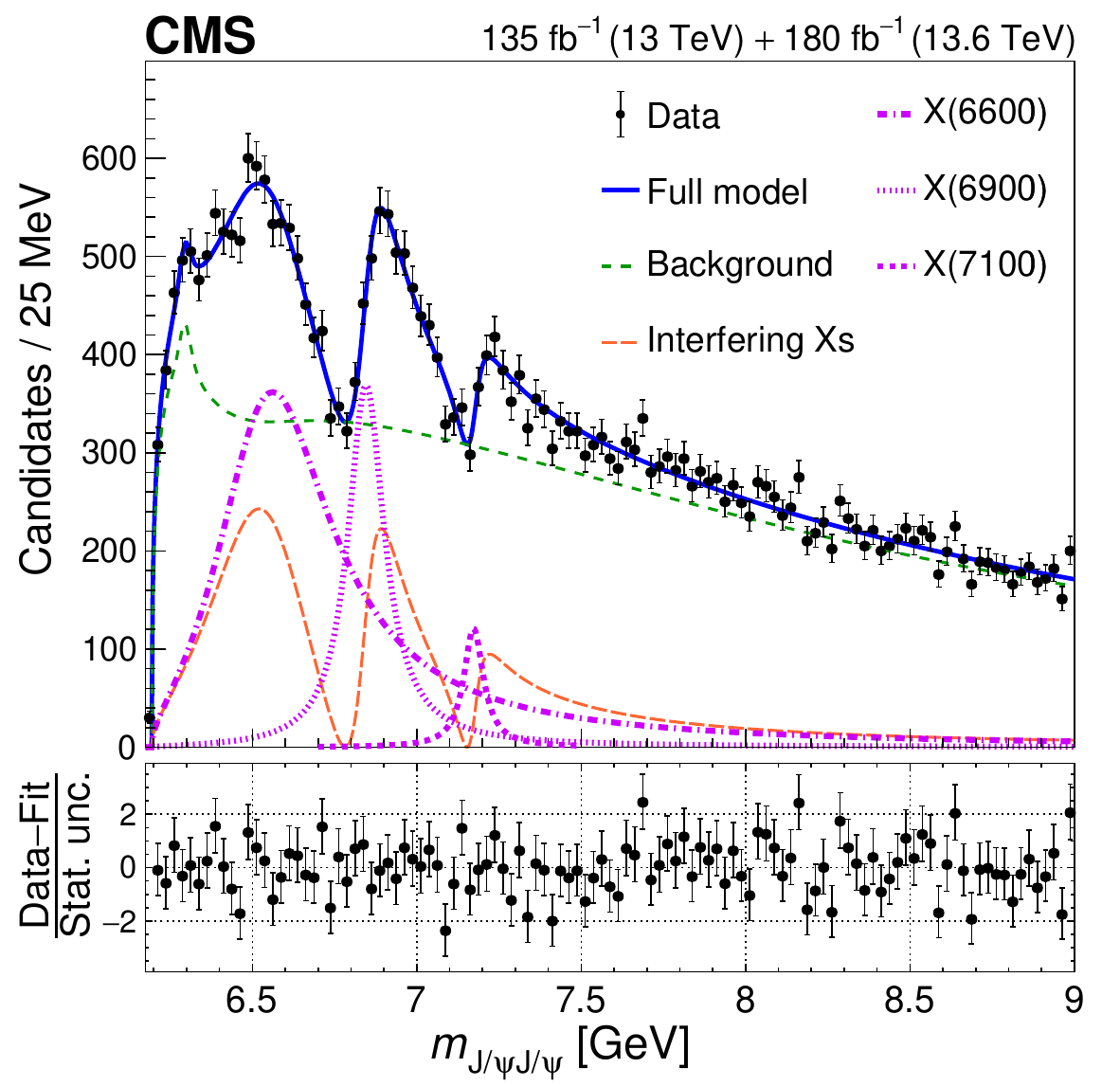}
\caption{
    \firstsentence{The \JJ invariant-mass spectrum up to 9\GeV.}
    The data is fit (up to 15\GeV, see text) with a three-way interference model using an unbinned likelihood, with the spectrum binned here for display.
    The model consists of three signal functions $[$\Xss, \Xsn, and $\Xso\!]$, and background components (NRSPS, DPS, combinatorial, a background threshold $\mathrm{BW}_0$, and $\PXP{6900} \to \JPp$ feed-down).
    The cumulative squared signal amplitude (proportional to $\abs{\mathcal{M}}^{2}$) is also shown (``Interfering Xs'').
    The lower panel shows the difference between the data and the fitted model, in units of standard deviations.
}
\label{Fig:Interf-fitmodel}
\end{figure}

Figure~\ref{Fig:Interf-fitmodel} shows the \JJ invariant mass spectrum up to 9\GeV, along with the results of the interference fit, which includes masses up to 15\GeV.  
In addition to the full fit, the combined background contribution, the individual BW signals, and the combined total squared amplitude of the interfering BW signals are shown.  
Results showing the individual background contributions and the full range of the fit are provided in the Supplementary Material. 
The results of the \JJ interference fit are given in Table~\ref{Tab:comparison_2_3_simul}.
Compared with the Run~2 result~\cite{cmscollaboration2023observation}, the three states show good consistency, with a notable \Xss mass shift, which is attributed to the addition of the feed-down component in the combined fit.
The addition of Run~3 data and the refined fit model improves the precision, reducing the statistical uncertainties in the BW parameters by about a factor of three in most cases.

The significances of these structures are established using a likelihood-ratio test~\cite{Wilks:1938dza}, where the likelihood of the nominal fit, including all components, is compared to that of a fit with the signal of interest omitted.
All structures have significances exceeding $5\sigma$.
The \Xso structure has a significance of $7.7\sigma$, surpassing the observation threshold of $5\sigma$ for the first time.
To test the robustness of the \Xso significance, we evaluated alternative fit-model hypotheses that explicitly incorporate the sources of systematic uncertainties described in Materials and Methods, none of which led to a substantial reduction in the significance.
Thus, the presence of the three distinct resonances is now established.

\begin{figure}[htb]
\centering
\includegraphics[width=0.49\textwidth]{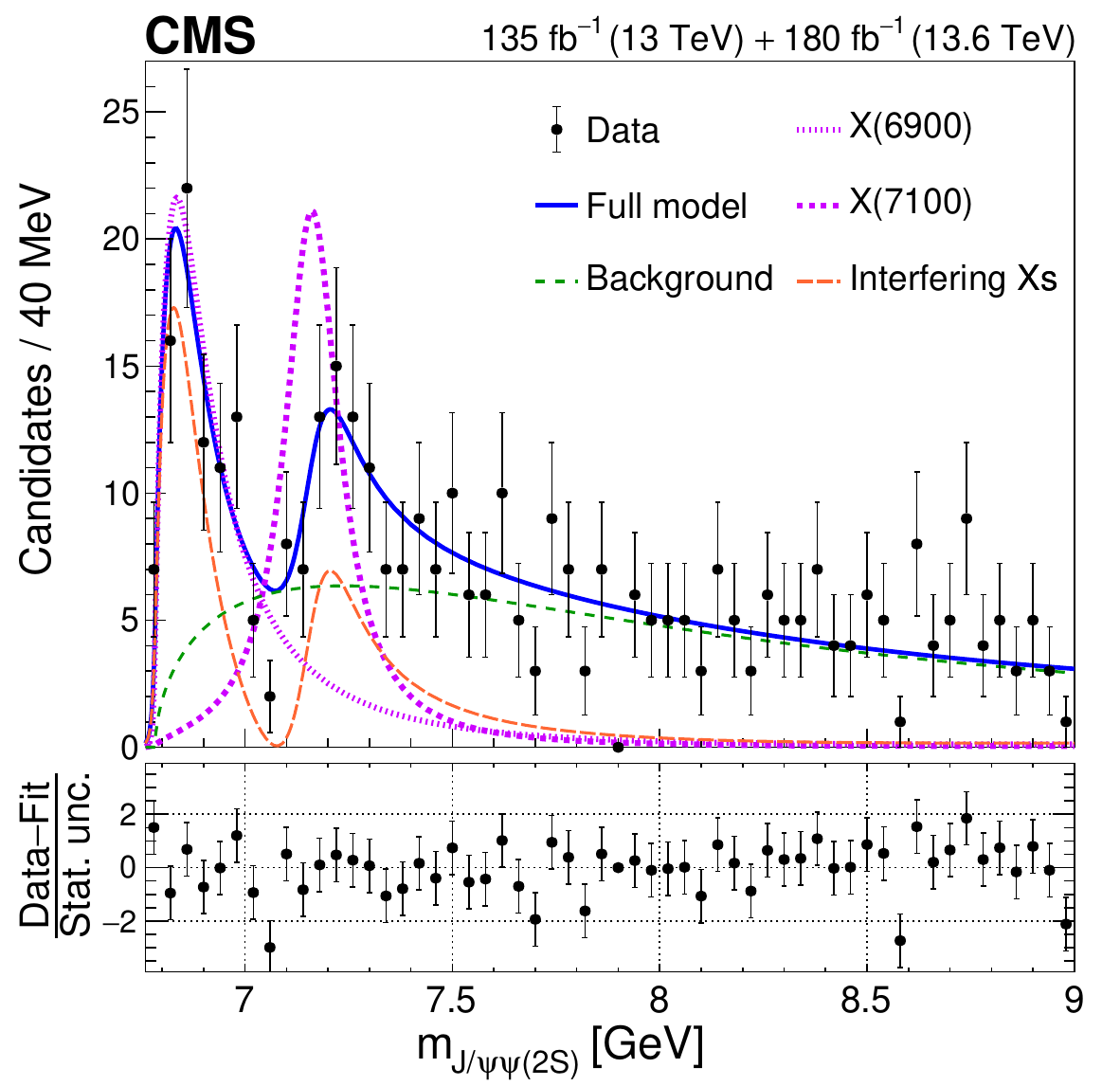}
\caption{
    \firstsentence{The \JPp invariant-mass spectrum up to 9\GeV.}
    The data is fit (up to 15\GeV, see text) with a two-way interference model, consisting of two signal functions [\Xsn and {\Xso}], and background components (NRSPS, DPS, and combinatorial).
    The cumulative squared signal amplitude is also shown (``Interfering Xs'').
    The lower panel shows the difference between the data and the fitted model, in units of standard deviations.
 }
\label{Fig:2way} 
\end{figure}

\begin{table*}[htb]
\topcaption
{
    \firstsentence{Measured masses and widths of the three \PX states from the fits to the \JJ and \JPp mass spectra from the Run~2+3 data set.} 
    The two uncertainties are statistical (first) and systematic (second).
    For comparison, the results from Run~2~\cite{cmscollaboration2023observation} are also shown.
    The values are in \MeVns.
}
\label{Tab:comparison_2_3_simul}
\centering
    \renewcommand\arraystretch{1.3} 
    \begin{tabular}{ c c c c c }
     & & \Xss & \Xsn & \Xso \\
    \hline            
    \multirow{2}{*}{\JJ: Run~2~\cite{cmscollaboration2023observation}} 
     & Mass  & $6638_{\,-\,38\,-\,31}^{\,+\,43\,+\,16}$ & $6847_{\,-\,28\,-\,20}^{\,+\,44\,+\,48}$ & $7134_{\,-\,25\,-\,15}^{\,+\,48\,+\,41}$ \\
     & Width & $440_{\,-\,200\,-\,240}^{\,+\,230\,+\,110}$ & $191_{\,-\,49\,-\,17}^{\,+\,66\,+\,25}$ & $97_{\,-\,29\,-\,26}^{\,+\,40\,+\,29}$ \\[4pt]
    \multirow{2}{*}{\JJ: Run~2+3} 
     & Mass  & $6593_{\,-\,14}^{\,+\,15} \pm 25$ & $6847 \pm 10 \pm 15$ & $7173_{\,-\,10}^{\,+\,9} \pm 13$ \\
     & Width & $446_{\,-\,54}^{\,+\,66} \pm 87$ & $135_{\,-\,14}^{\,+\,16} \pm 14$ & $73_{\,-\,15}^{\,+\,18} \pm 10$ \\[4pt]
    \multirow{2}{*}{\JPp: Run~2+3} 
     & Mass  & \NA & $6876_{\,-\,29\,-\,112}^{\,+\,46\,+\,111}$ & $7169_{\,-\,52\,-\,72}^{\,+\,26\,+\,80}$ \\
     & Width & \NA & $253_{\,-\,100\,-\,134}^{\,+\,290\,+\,140}$ & $154_{\,-\,82\phantom{1}\,-\,181}^{\,+\,110\,+\,140}$ \\
    \end{tabular}

\end{table*}

For the \JPp events, only two BW terms, \Xsn and \Xso, are included in the default mass distribution fit model, as the mass of the \Xss state is below the kinematic threshold. 
Given that the \Xsn signal is near threshold, where the sensitivity to resolution and efficiency effects is enhanced, these effects are incorporated in the default \JPp model by convolving the signal shapes with the resolution functions and applying corrections for efficiency variations across the entire resonance region.
The \JPp background model consists of three sources: NRSPS, DPS, and combinatorial background.
The modeling of the three background components follows that used for the \JJ events.

The \JPp invariant mass spectrum is shown in Fig.~\ref{Fig:2way}, along with the results of the likelihood fit including the sum of all components, the two individual BW signals, the combined signal (including interference), and the sum of the three background contributions.
The $\chi^2/\nbin$ of the fit, with \nbin being the number of nonzero bins, calculated using the bins of Fig.~\ref{Fig:2way} and with masses below 9\GeV, is 57/56, indicating a good description of the data.
The resulting masses and widths of the two peaks are given in Table~\ref{Tab:comparison_2_3_simul}.
Destructive interference between the \Xsn and \Xso structures is again apparent.
An alternative hypothesis where the \Xsn contribution is replaced by the tail of the \Xss also provides a similarly good fit to the data with $\chi^2/\nbin = 57/56$ for bins below 9\GeV. 
However, the clear evidence for the \Xsn state in the \JJ events provides a compelling motivation to include both the \Xsn and the \Xso in the description of the spectrum.
Furthermore, the mass and width of the \Xsn from the \JPp events are consistent with those from our \JJ data (Table~\ref{Tab:comparison_2_3_simul}), supporting the interpretation that the excess of \JPp candidates above the background-only model for masses below 7\GeV is the \Xsn structure.
Although contributions to the excess from threshold effects~\cite{Dong:2020,Guo:2019twa,Gong:2020bmg} cannot be excluded with the amount of data available, the concordance between the \JJ and \JPp results suggests that other processes do not have a large effect.

Because the \JPp sample is significantly smaller, the \JJ events provide more precise measurements; therefore, when performing the likelihood ratio test with the \JPp events to estimate the significances of \Xsn and \Xso, the masses and widths of both structures are constrained to lie within one standard deviation (with statistical and systematic uncertainties added in quadrature) of the values obtained from the \JJ fit.
When comparing the likelihood of the nominal fit to that of fits with each signal, \Xsn or \Xso, omitted, we find their significances to be $8.1\sigma$ and $4.3\sigma$, respectively.

All three LHC experiments favor models incorporating some sort of interference~\cite{cmscollaboration2023observation,LHCb:2020bwg,ATLAS:2023bft}, but none report a conclusive statistical significance.
The interference fit to the \JJ events returns a $\chi^2/\nbin = 52/65$ in the signal region below 7.8\GeV, compared with 147/65 for the noninterference fit.
To quantify the role of interference, we perform alternate fits omitting specific interference terms and compare the resulting log-likelihoods with that of the full interference model fit. 
The dips near 6750 and 7150\MeV correspond to the \Xss--\Xsn and \Xsn--\Xso interference terms.
Omitting each term worsens the description of the data, and the significances of the dips, evaluated with a likelihood-ratio test relative to the nominal fit, are $9.7\sigma$ and $6.5\sigma$, respectively.
For the \JPp events, the significance of the \Xsn--\Xso interference dip from the likelihood ratios is $2.5\sigma$.

\section{\scihead{Discussion}} 
\subsection{Patterns within the \texorpdfstring{\JJ}{J/psi J/psi} family} 
Particles are normally characterized by their total angular momentum ($J$), parity ($P$), and charge conjugation ($C$) quantum numbers, abbreviated as $J^{PC}$.
Numerous spin-parity states are allowed for tetracharm states ($J^{PC}$ = $0^{++}$, $0^{+-}$, $0^{-+}$, $1^{++}$, $1^{+-}$, $1^{-+}$, $2^{++}$, \ldots).
However, the \JJ decay mode dictates that $C_{\PX} = (-1)(-1) = +1$ because $J^{PC}_{\Jpsi} = 1^{--}$ and $C$ is conserved in strong decays.
Furthermore, quantum interference among the \PX structures implies that the states are coherently produced and share a common $J^{PC}$.
Particles with the same quark content and same $J^{PC}$ quantum numbers can have different masses as a result of different internal orbital ($L$) or spin ($S$) angular momentum quantum numbers or from radial excitations, which are characterized by the quantum number $n$.

The large \PX-mass splittings ($>$250\MeV) are characteristic of radial excitations, as illustrated by the $\PGUP{nS}$ bottomonium family~\cite{PhysRevD.110.030001}, which consists of spin-1 $\PQb\PAQb$ states in the S-wave ($L=0$) configuration and represents the $\PQb\PAQb$ counterpart of the \Jpsi family.
In contrast, the mass splittings resulting from different values of $L$ and $S$ are much smaller.
Studies of our Run~2 results found that the pattern of \PX masses may be indicative of the radial variant of a Regge trajectory~\cite{FengZ}, \ie, $n = \beta m^2 + \beta_0$, where $n$ is the radial number, $m$ is the mass of the state, and $\beta$ and $\beta_0$ are constants. 

\begin{figure}[htb]
\centering
\includegraphics[width=0.49\textwidth]{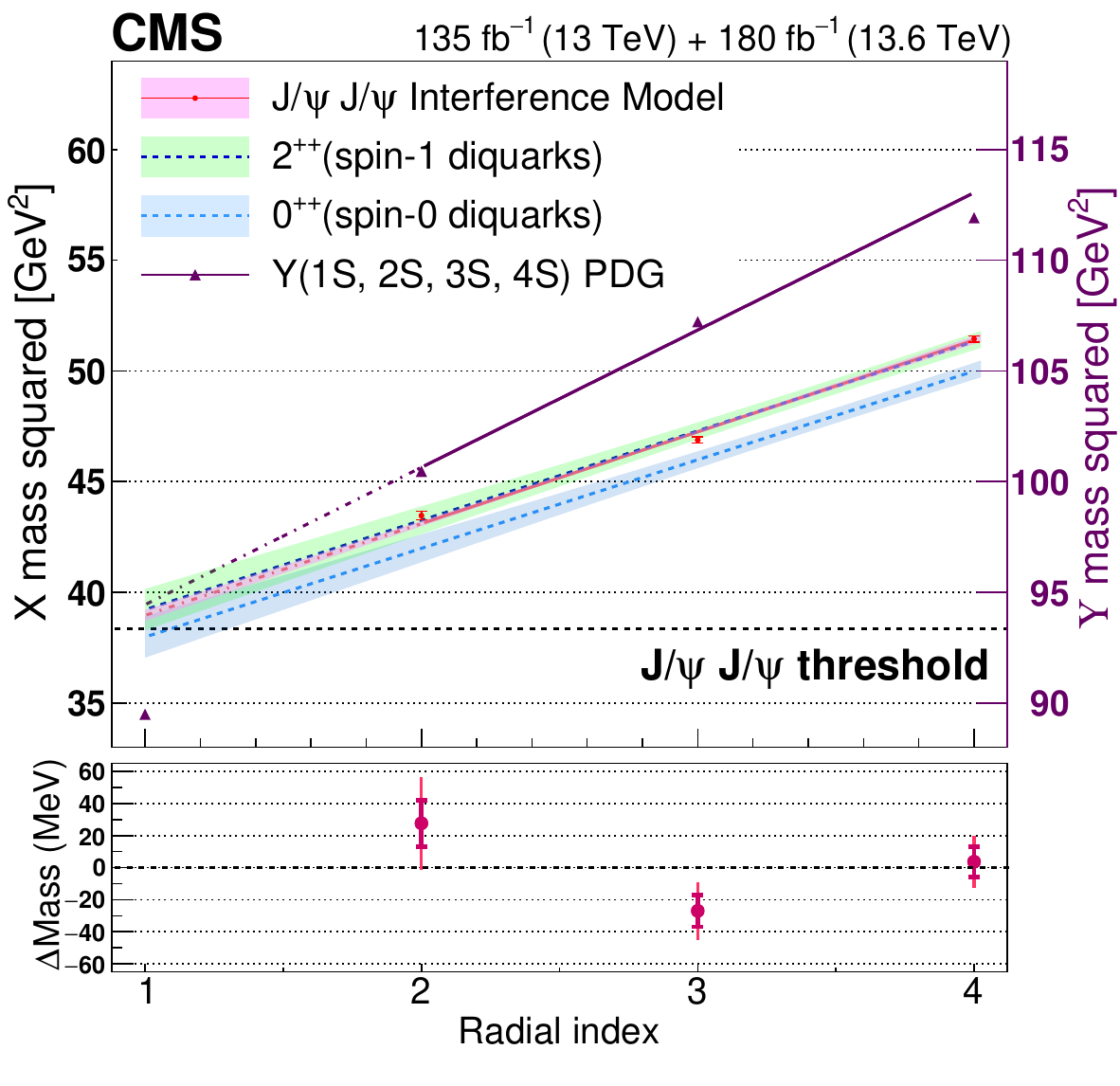}
\caption
{
    \firstsentence{Squared masses of \Xsn, \Xss, \Xso (statistical  uncertainties only), and Upsilon (\PGU, masses from Ref.~\cite{PhysRevD.110.030001}, denoted as ``PDG''), families versus radial indices $n=2$, 3, 4.}
    The solid lines are linear fits for the \PX and \PGU families (see text), with dashed extrapolations to projected $n=1$ states.
    Regge trajectories calculated for $0^{++}$ (spin-0 diquarks) and $2^{++}$ (spin-1 diquarks) tetraquarks are also shown.
    Bands correspond to $\pm$1$\sigma$ statistical uncertainties for data (pink band) and the fully correlated $\pm$1$\sigma$ uncertainty from the strong coupling constant $\alpha_s$ for theory (green and blue bands)~\cite{Zhu:2020xni}. 
    The lower panel shows the deviations of the measured $\PX$ masses from their fitted trajectory with error bars showing the statistical and total uncertainties.
}
\label{Regge}
\end{figure}

To determine the Regge trajectory, we fit the four-muon invariant mass out to 15\GeV, replacing the BW mean parameters for \Xss, \Xsn, \Xso with $m_n = \sqrt{\smash[b]{(n-\beta_0)/\beta}}$ $(n=2,3,4)$ where $\beta$ and $\beta_0$ are free parameters. 
The index choice is based on assigning $n=3$ to the \Xsn state, as done in Ref.~\cite{Zhu:2020xni}.  
It should be noted, however, that any other sequential indexing, such as $n=1,2,3$, will produce equivalent results (with a different $\beta_0$ value).  
The squared measured masses from the original fit and the fitted Regge trajectory are shown in Fig.~\ref{Regge}, with the mass deviations between the two results shown in the lower panel.  
Based on the match of the data to a Regge trajectory and the shared $J^{PC}$ values inferred from the interference, we deduce that the \PX states comprise a family of radial excitations---$\ie$, a set of states with the same quantum numbers except for the radial one, and likely with a common underlying physical configuration.

The squared $\PGUP{nS}$ masses for $n=1$--$4$ are also shown, along with a linear fit to the $n=2$--$4$ results. 
The $n=1$ mass is dropped from the fit because it falls significantly below the Regge trajectory of the higher states. 
Ground states are often found to have masses lighter than the extrapolation from their excited states~\cite{Ebert:ReggeHeavy}, which can be understood from the structure of the interquark potential. 
Heavy quarkonia are commonly described by the Cornell potential~\cite{Ebert:ReggeHeavy}, which consists of a short-range Coulomb-like term and a long-range linear confinement term: $V(r) = -(4/3)\,\alpha_s/r + \lambda\,r$, where $r$ is the radial distance, $\alpha_s$ is the strong coupling constant, and $\lambda$ denotes the string tension.
The ground states often deviate from linear trajectories because their constituents occupy the region dominated by the Coulomb part of the potential rather than the linear confinement region.

Various theoretical models predict Regge trajectories for all-charm tetraquarks.
Figure~\ref{Regge} shows the Regge trajectories from Ref.~\cite{Zhu:2020xni}, where particle masses up to $n=4$ were calculated using a Bethe--Salpeter analysis of either spin-0 or spin-1 diquarks in an S-wave configuration within a Cornell potential supplemented by spin-interaction terms.
As noted earlier for the $\PGUP{nS}$ states, ground states often deviate from linear trajectories.
We therefore refit the Regge trajectories from Ref.~\cite{Zhu:2020xni}, dropping the $n=1$ states while using their calculated masses. 
The refitted theoretical trajectories are shown in Fig.~\ref{Regge} for $0^{++}$ states from spin-0 (color sextet) diquarks and $2^{++}$ states from spin-1 (color triplet) diquarks.
The $0^{++}$ trajectory from spin-1 diquarks is not shown, as it overlaps the $2^{++}$ result.
Both theoretical results show a similar slope to our data, with the spin-1 diquark result being nearly indistinguishable from the data and the spin-0 diquark result lying slightly below. 
Comparisons with other theoretical calculations are provided in the Supplementary Material, including other calculations of the same $L=0$ orbital angular momentum and with $L=1$ and $2$ configurations.
The best agreement with the data is obtained using the spin-1 diquark model of Ref.~\cite{Zhu:2020xni}, shown in Fig.~\ref{Regge}. 
Nevertheless, theoretical predictions exhibit sizable variations even within the same configuration, leaving room for alternative interpretations.

\begin{figure}[htb]
\centering
\includegraphics[width=0.49\textwidth]{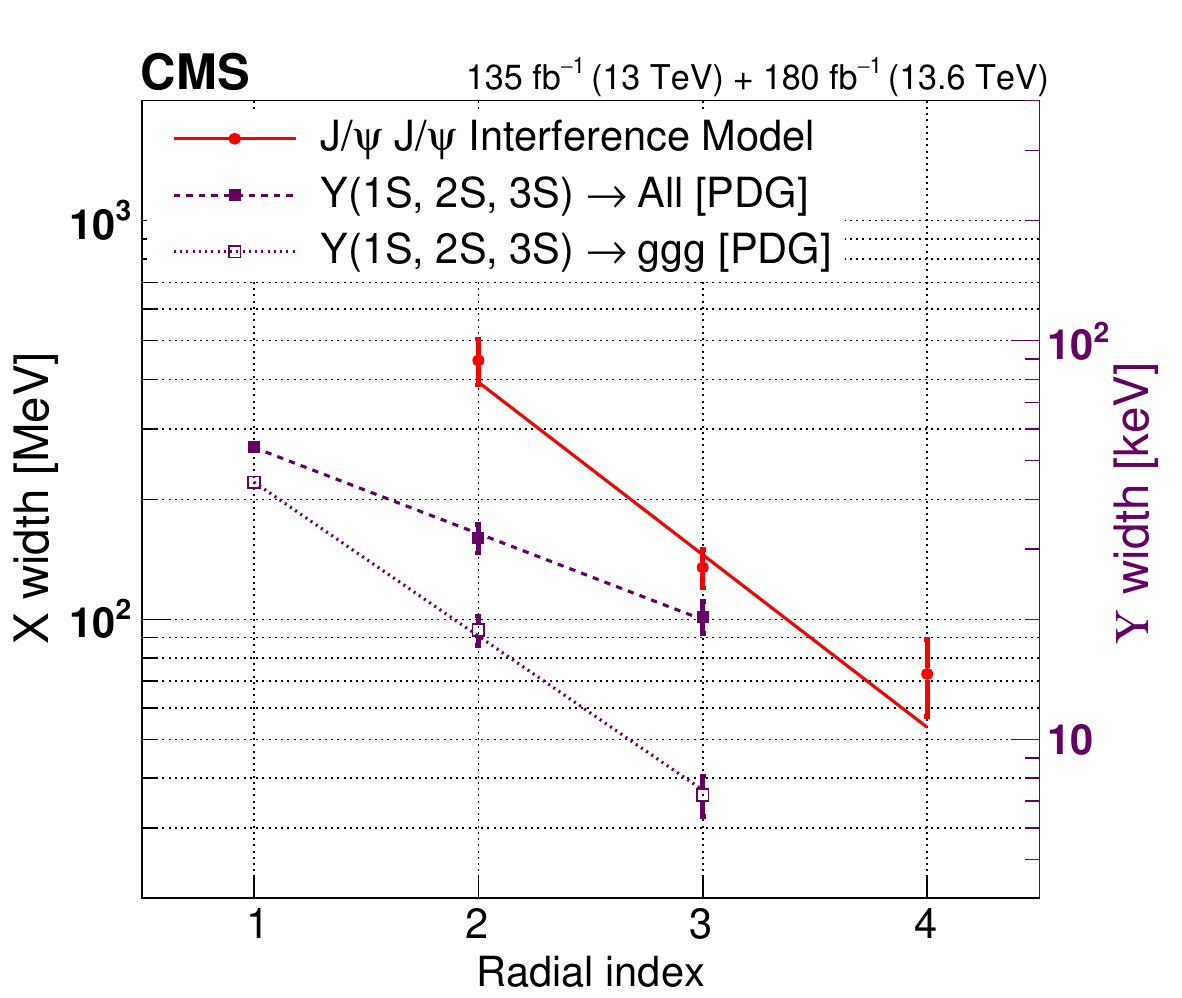}
\caption{ 
\firstsentence{The widths of the \PX states as a function of the radial quantum number (statistical uncertainties only).} 
For comparison, the total widths and the three-gluon ($3\Pg$) partial widths of the \PGUP{1S}, \PGUP{2S}, and \PGUP{3S} states are also displayed~\cite{PhysRevD.110.030001}, denoted as ``PDG''.
The lines are fits to exponential functions.
}
\label{Gamma}
\end{figure}

Another parameter characterizing resonances is their natural decay width, $\Gamma$. 
The widths of the \PX states systematically decrease with radial excitation, as shown in Fig.~\ref{Gamma}. 
The trend can be described by an exponential function, with a fitted ratio $\Gamma_\PX(n+1)/\Gamma_\PX(n) = 0.37 \pm 0.07$, having a $p$-value of 7.5\%, considering only statistical uncertainties.

The natural widths, which are inversely proportional to a particle's lifetime, are often expected to increase with mass because the available kinematic phase space of the decays grows, and more decay channels open up.
A counter example whose behavior is similar to the \PX widths is the first three members of the \PGUP{nS} family, which is illustrated in Fig.~\ref{Gamma}.
As these \PGU states are below the 10.559\GeV threshold for decays to pairs of $\PQb$ mesons, decays that would otherwise be small can play a prominent role.
One such process is $\PQb\PAQb$ annihilation via three gluons (3\Pg), which occurs in $\sim$82\% of $\PGUP{1S}$ decays~\cite{PhysRevD.110.030001}. 
The 3\Pg partial widths also exponentially decrease (Fig.~\ref{Gamma}), and a fit yields $\Gamma_{\PGU\to 3\Pg}(n+1)/\Gamma_{\PGU\to 3\Pg}(n) = 0.41 \pm 0.02$.
The quantitative similarity of the two width ratios may be coincidental, but their similarity is suggestive of analogous decay mechanisms.  
The decrease of $\Gamma_{\PGU\to 3\Pg}$ is understood to occur because annihilation requires the quarks to come into ``contact'', and this probability is proportional to $\abs{\varphi_n(0)}^2 / M_n^2$ in a simplified nonrelativistic model~\cite{PhysRevD.17.3090}, where $\abs{\varphi_n(0)}$ is the magnitude of the \PGUP{nS} state's wave function for quark separation $r=0$, and $M_n$ is its mass.
As the radial index of the $\PQb\PAQb$ system increases, the wave function expands spatially and $\abs{\varphi_n(0)}$ decreases, leading to a decreasing partial width~\cite{Ke:2010UpsWidths}.

We measure the full widths of the $\PX$ states, which are the sum of the partial widths of all their decay modes. 
The modes producing mesons may be categorized, in their simplest form, as radiative, annihilation, or rearrangement/fall-apart decays, all of which proceed predominately via the strong interaction. 
Radiative decays are where the tetraquark system radiates gluons, and the gluons hadronize into light hadrons (\eg, $\PX+\text{(light hadrons)}$ and $\PJGy\PJGy +\text{(light hadrons)}$). 
In annihilation decays one, or both, of the $\PQc\PAQc$ pairs annihilate into $\PQq\PAQq$ pairs producing the minimal quark content, respectively, of $\PQc\PAQc\PQq\PAQq$ (\eg, $\PD^{(*)}_{(s)}\PAD{}^{(*)}_{(s)}$ and $\PJGy+$(light hadrons)), or $\PQq\PAQq\PQq'\PAQq'$ (\ie, light hadrons). 
Finally, there are processes where the quark content is unchanged (e.g., $\PGhc\PGhc$ and $\PJGy\PJGy$).
In a diquark model, these are rearrangement decays in which each charm quark from the diquark matches with a charm antiquark from the antidiquark to produce pairs of charmonium mesons, and in a molecular model, the molecule simply falls apart into two $\PQc\PAQc$ mesons.
 
In a diquark model, annihilation decays, such as $\PX {\to} \PD^{(*)}_{(s)}\PAD{}^{(*)}_{(s)}$, are similar to $\PGU \to 3\Pg$ in that the decay width will depend on the degree of overlap of diquark wave functions, which should lead to decreasing partial widths as $n$ increases.
Rearrangement decays, like $\PJGy\PJGy$, do not involve the localization of quarks required for annihilation, but their probability should again depend on the overlap of diquark wave functions. 
Theoretical calculations for $0^{++}$ and $2^{++}$ tetraquarks in a diquark model support the expectation that the diquark wave function at the origin is smaller for 2S states than 1S   states~\cite{Debastiani:DiquarkSize2017,2409.12070}.  
 
To the extent that rearrangement and annihilation decays, like $\PX {\to} \JJ$ and $\PD^{(*)}_{(s)}\PAD{}^{(*)}_{(s)}$, are a large fraction of the full \PX widths, a diquark model offers a promising account for the observed pattern of \PX widths, but quantitative consensus among estimations of tetraquark wave functions is lacking~\cite{Debastiani:DiquarkSize2017,2409.12070}.

\subsubsection{Additional family members}
The Regge trajectory of the \PX states allows us to project where other family members---if they exist---should lie.
If our radial indexing is correct, and the trajectory is valid near threshold, 
the dot-dashed extrapolation in Fig.~\ref{Regge} places the $n=1$ state around 6240\MeV.
However, as is also shown in Fig.~\ref{Regge}, the \PGUP{1S} state lies significantly below the $n=1$ position inferred from the linear fit. 
By analogy, the $n=1$ state in the \JJ family may also be expected to fall below the extrapolated prediction~\cite{Zhu:2020xni},
although not by as much as the \PGUP{1S}, which is tightly bound in the Coulombic part of the potential~\cite{Ebert:ReggeHeavy}.
In this mass region, the ATLAS, CMS, and LHCb experiments all find a near-threshold excess, which remains uninterpreted~\cite{cmscollaboration2023observation,LHCb:2020bwg,ATLAS:2023bft}.
There are predictions for a state that could explain this excess, possibly below the \JJ invariant-mass threshold~\cite{Zhu:2020xni,Debastiani:DiquarkSize2017,Song:2024ykq}.
Our data may be able to accommodate the presence of the $n=1$ state in the \JJ family but, if present, it is obscured by the feed-down events (illustrated in the Supplementary Fig.~\insupp{S1}{\ref{Fig:full_range}}), and a better understanding of the threshold region is necessary to draw a conclusion about a potential fourth family member.

Projecting the Regge trajectory of the \PX states indicates that the next higher family member, if there is one, should be at a mass of $7453 \pm 21$\MeV---but the data offer no visual hint of such a state (Fig.~\ref{Fig:Interf-fitmodel}).
The presence of this potential signal was checked in our data for a wide range of parameters, with no significant indication found. 
The apparent absence of a \JJ signal for such a state is not surprising in a diquark model, even if this family member exists, as this mass is well above the $\PGXpcc \PAGXmcc$ and $\PGXppcc \PAGXmmcc$ thresholds ($\sim$7240~MeV~\cite{LHCb:2021Xiccp,LHCb:2022Xiccpp}); and a highly excited [$\PQc\PQc$][$\PAQc\PAQc$] system can pull a $\PQq\PAQq$ pair from the vacuum and readily decompose into [$\PQc\PQc\PQq$][$\PAQq\PAQc\PAQc$], \ie a $\PGXcc\PAGXcc$ pair.
This would suppress the branching fraction to \JJ along with increasing its width. 
In contrast, in a molecular model, a [$\PQc\PAQc$][$\PQc\PAQc$] system has to undergo the additional step of quark rearrangement for baryon decays, and thus the \JJ decay might be a substantial decay mode.

\subsection{Interpretation and conclusions}
With almost four times more \JJ pairs than previously available~\cite{cmscollaboration2023observation}, the data are well described by mutually interfering \Xss, \Xsn, and \Xso states.
The statistical uncertainties in the masses and widths are reduced by a factor of three, benefiting not only from the increased number of events, but also from the improved fit model, and all three structures are established with significances well above 5$\sigma$ for the first time.
In the $\JPp \to \PGmp \PGmm \PGmp \PGmm$ events,
the \Xsn and \Xso structures are seen with significances exceeding $8\sigma$ and $4\sigma$, respectively.
In the \JJ events, interference among all three states is 
statistically compelling $({>}5\sigma)$. 
The \JPp result corroborates interference between \Xsn and \Xso at the $2.5\sigma$ significance level.
Interference implies common $J^{PC}$ quantum numbers for the three states.
This is further supported and clarified by the \PX masses displaying (Fig.~\ref{Regge}) radial Regge behavior, suggesting that the triplet represents a family of radial excitations of a common underlying configuration.

An interpretation of the triplet of \PX states based on a configuration of [$\PQc\PQc$][$\PAQc\PAQc$] diquarks seems to account for the pattern of masses, consistent with a Regge trajectory, and with the decreasing natural widths, consistent with decays that are dominated by annihilation or rearrangement processes. 
In contrast, these features are problematic for molecular or threshold interpretations.
In either of these cases, \JJ structures would be governed by the underlying charmonium pairs generating the structure [\eg,\ $\Jpsi\,\PGyP{3770}$ for the {\Xsn}].
There are numerous charmonia pairings possible in the region of the triplet, but charmonia are generally irregularly spaced, so that the linear Regge behavior is not expected.
Similarly, systematic trends among widths, let alone an exponentially decreasing one, are uncharacteristic of these interpretations, because different pairings are not expected to have simple correlations among their widths.
However, this does not necessarily preclude the presence of molecular subcomponents~\cite{LHCb:2020-3872MolecCompact,He:2024-Zb-DiMolec}, or a resonant interplay with threshold effects~\cite{Bugg:2008Cusp}.
Diquark and molecular models are idealized extremes of a continuum of possible internal configurations of four-quark systems.
Another theoretical configuration is a four-body system with no systematic quark clustering. 
Expectations of any mass or width trends of such ``amorphous'' systems require detailed calculations and, without theoretical guidance, the possibility of an amorphous structure for the \PX triplet cannot be excluded.  
While hybrid models have predicted low-lying $\PQc\PQc\PAQc\PAQc\Pg$ states in the vicinity of the \Xsn (for $0^{++}$) and \Xso (for $0^{-+}$), the different quantum numbers for the states cannot explain the observed interference and there is no associated \Xss state~\cite{Wan:2020fsk,tang2025}.

The Regge trajectory of the \PX states is found to match the predictions for a pair of spin-1 diquarks with \PX $J^{PC}$ quantum numbers of $0^{++}$ or $2^{++}$.
A recent CMS angular analysis of $\PX \to \JJ$ decays, assuming the same $J^{PC}$ for the three states, strongly favors $J^{PC} = 2^{++}$ over other potential quantum numbers~\cite{cms:SpinParity}.
In addition, production of a $0^{++}$ state is predicted to be suppressed compared to a $2^{++}$ state~\cite{Zhu:2020xni,Ma:2020kwb,Becchi:2020uvq}.
Combining this information leads to a consistent picture of a system of aligned spin-1 diquarks with no orbital angular momentum ($L_{[\PQc\PQc][\PAQc\PAQc]}=0$).
The observed structures would therefore be denoted as a triplet of $\mathrm{T}_{\PQc\PQc\PAQc\PAQc}$ states.
In the $n\, ^{2S+1}L_J$ spectroscopic notation, the simplest scenario is that they are radially excited  $n\, ^5 S_2$ states (``S'' indicating $L=0$) 
where, because our data does not discriminate between them, $n=1, 2, 3$ or $2, 3, 4$.
Other configurations with 
$L_{[\PQc\PQc][\PAQc\PAQc]}\geq 2$ are possible,
but high orbital excitations are more difficult to produce, and thus less likely.
Although the hypothesis of a triplet of $2^{++}$ states made up of spin-1 diquarks describes our data well, the picture is potentially more complicated because other tetraquark states may contribute to the \JJ spectrum (\eg, $0^{++}$~\cite{Zhu:2020xni,Ma:2020kwb,Tiwari:2021tmz});
or members of the triplet may consist of mixtures of different internal configurations with the same $J^{PC}$, such as pairs of spin-0 and anti-aligned spin-1 diquarks~\cite{Wang:2021kfv}, or include a molecular subcomponent~\cite{LHCb:2020-3872MolecCompact,He:2024-Zb-DiMolec}.
Further exploration, experimental and theoretical, of all-heavy tetraquarks is required to resolve these more subtle issues.

\section{\scihead{Materials and methods}}
\subsection{Experimental data}
The events used in this analysis were recorded by the CMS detector~\cite{CMS:2008xjf,Hayrapetyan_2024}, which is a general-purpose experiment optimized for high-mass processes such as those involving the \PW, \PZ, and Higgs bosons, while still retaining a significant capability for investigating other phenomena like heavy-quark hadron spectroscopy.  
CMS uses a right-handed coordinate system with the $x$ axis pointing to the center of the LHC, $y$ pointing up, and $z$ along the beam axis.  
Particle momenta are typically characterized by the pseudorapidity $\eta = -\ln[\tan(\theta/2)]$, where $\theta$ is the polar angle relative to the beam axis, and the transverse momentum $\pt$, which is the component of the momentum perpendicular to the beam axis.  
The detector includes a silicon pixel and strip tracker, a lead tungstate electromagnetic calorimeter, and a brass and scintillator hadron calorimeter, each with barrel and endcap sections, all of which are located inside a 6-meter-diameter, 13-meter-long, superconducting solenoid generating a 3.8\unit{T} magnetic field.  
Outside of the solenoid is the steel flux-return yoke, interspersed with gas-ionization muon detectors in a 2\unit{T} field.  
Nearly all particles are absorbed by the calorimeters, leaving muons as the only charged particle to reliably pass through all these detectors.  
Charged particles are reconstructed from hits in the silicon tracker and identified as muons by matching these tracks with hits in the muon detectors, which cover $\abs{\eta}<2.4$.

At the LHC, only a small fraction of $\Pp\Pp$ collisions produced can be recorded.
Events of interest are selected using a two-tiered trigger system~\cite{CMS:2016ngn}. 
The Level-1 (L1) trigger, implemented in custom electronics, reduces the input event rate using partial detector information.
The high-level trigger (HLT)~\cite{CMS:2024aqx}, which is software running on commercial processors, performs a full event reconstruction to further reduce the rate to a few kilohertz for permanent storage.

To select $\JJ$ and $\JPp$ candidate events, two sets of trigger conditions are used.
For Run~2 data from 2016--2018, the trigger requires at least three muons at L1. In 2017--2018, \pt requirements were added to ensure at least two muons have $\pt > 3\GeV$ and at least one with $\pt > 5\GeV$.
At the HLT, events are required to contain an oppositely charged muon pair with an invariant mass between 2.95 and 3.25\GeV, a distance of closest approach between the two muons less than 0.5\cm, and a successful vertex fit with a $\chi^2$ probability greater than 0.5\%.
For the 2017--2018 data, an additional requirement was applied that each muon in the pair has $\pt >3.5\GeV$.
In Run~3, the introduction of the parking stream~\cite{CMS:2024zhe} enabled the use of an additional trigger.
At L1, at least two muons are required.
The HLT selects oppositely charged muon pairs with an invariant mass below 8.5\GeV and $\pt > 4.9\GeV$, where both muons must satisfy $\pt>3\GeV$ and one must have $\pt>4\GeV$.
For the Run~3 data, both trigger paths described above are employed, combined using a logical OR condition.

For the final \JJ event selection, we adhere to the criteria detailed in Ref.~\cite{cmscollaboration2023observation}. 
There must be at least four muons with $\pt>2\GeV$, $\abs{\eta}<2.4$, and satisfying the ``soft muon identification'' requirements~\cite{CMS:2012nsv}.
Muon pairs of opposite charge are used to form \Jpsi candidates, required to have $\pt > 3.5\GeV$, and a fit probability greater than 0.5\% when fitting the two muons to a common vertex.
The muons are ordered by the \pt of the reconstructed \Jpsi candidates, with $\PGm_1^{+}\PGm_2^{-}$ forming the $\Jpsi_1$ with higher \pt, and thus of higher purity.
A mass-constrained vertex fit is applied to muon pairs with opposite charge, in which the muon pair's mass is constrained to the known \Jpsi mass of 3096.9\MeV~\cite{PhysRevD.110.030001}, and the fit probability is required to be greater than 0.1\%.
In some events, multiple \JJ candidates can be formed from different charge permutations of the same four muons.
If both charge permutations of the same four muons have masses within $3\sigma_m$ of the world-average \Jpsi mass~\cite{PhysRevD.110.030001}, where $\sigma_m$ is the calculated uncertainty of the reconstructed dimuon mass, we select the pairing that minimizes $(\Delta m_1/\sigma_{m_1})^2+(\Delta m_2/\sigma_{m_2})^2$, where $\Delta m$ is the difference between the reconstructed dimuon mass and the nominal \Jpsi mass.
All \JJ candidates are kept if they arise from more than four muons, which occurs in 0.2\% (0.4\%) of Run~2 (Run~3) events.

The \JPp selection was patterned on the \JJ selection, but the $\Pps \to \Pgmp \Pgmm$ signal suffers from larger backgrounds and has a smaller branching fraction than for  $\Jpsi \to \Pgmp \Pgmm$, thus requiring stricter criteria.
A selection optimization is performed to maximize the signal significance using 2016-18 data and the resulting \JPp selection is as follows.
There must be four muons with zero net charge and satisfying: $\pt(\Jpsi) >  11\GeV$; $\pt(\Pps) > 13.5\GeV$; each \Pps muon satisfies the ``loose muon identification''~\cite{CMS:2012nsv} criteria and has $\pt > 2.5\GeV$; and the dimuons must fall within $2.5\sigma_m$ of the known \Pps ($\PGm_1\PGm_2$) or \Jpsi ($\PGm_3\PGm_4$) masses~\cite{PhysRevD.110.030001}, roughly [3610, 3760]\MeV or [3020, 3180]\MeV, respectively.
Finally, the four muons are fit to a common vertex, with opposite-sign pairs constrained to either the \Pps or \Jpsi masses, and kept if the fit probability exceeds 0.5\%.

\begin{figure*}[htbp]
\centering
\includegraphics[width=0.45\textwidth]{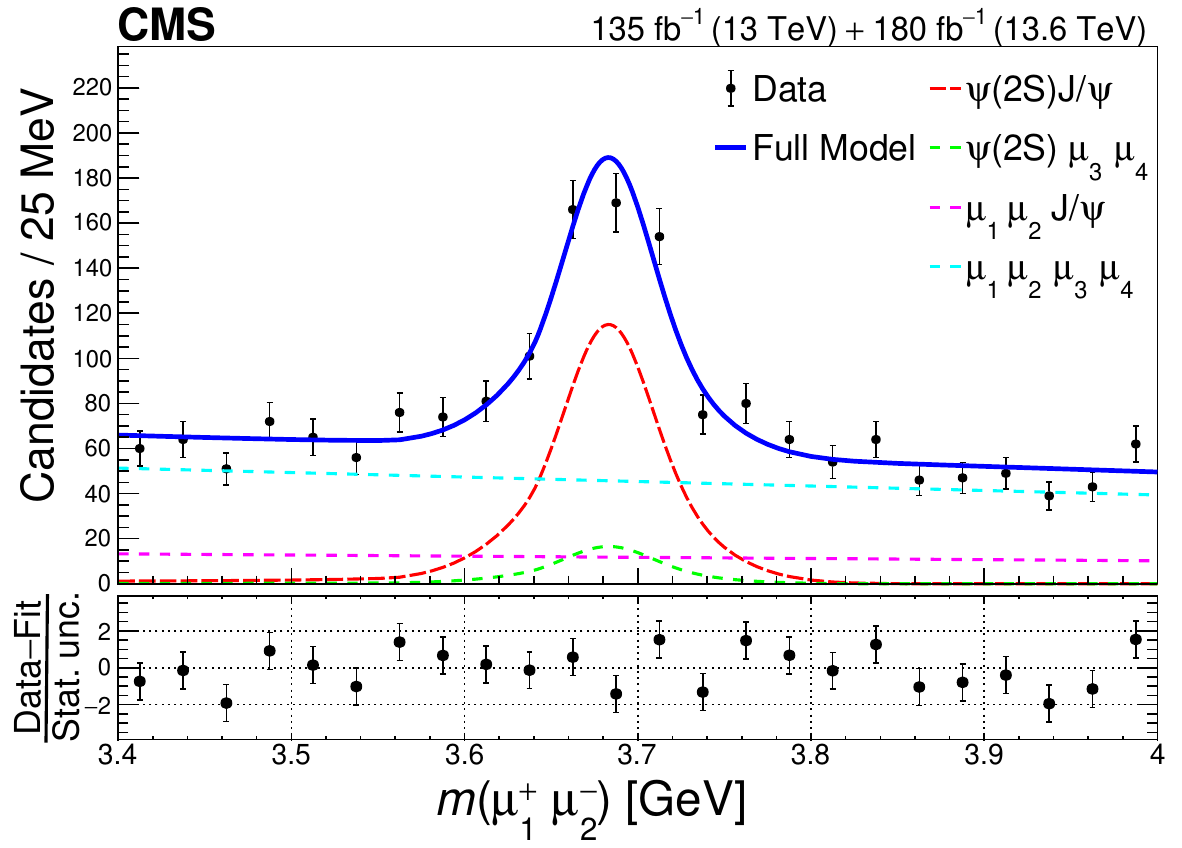}
\includegraphics[width=0.45\textwidth]{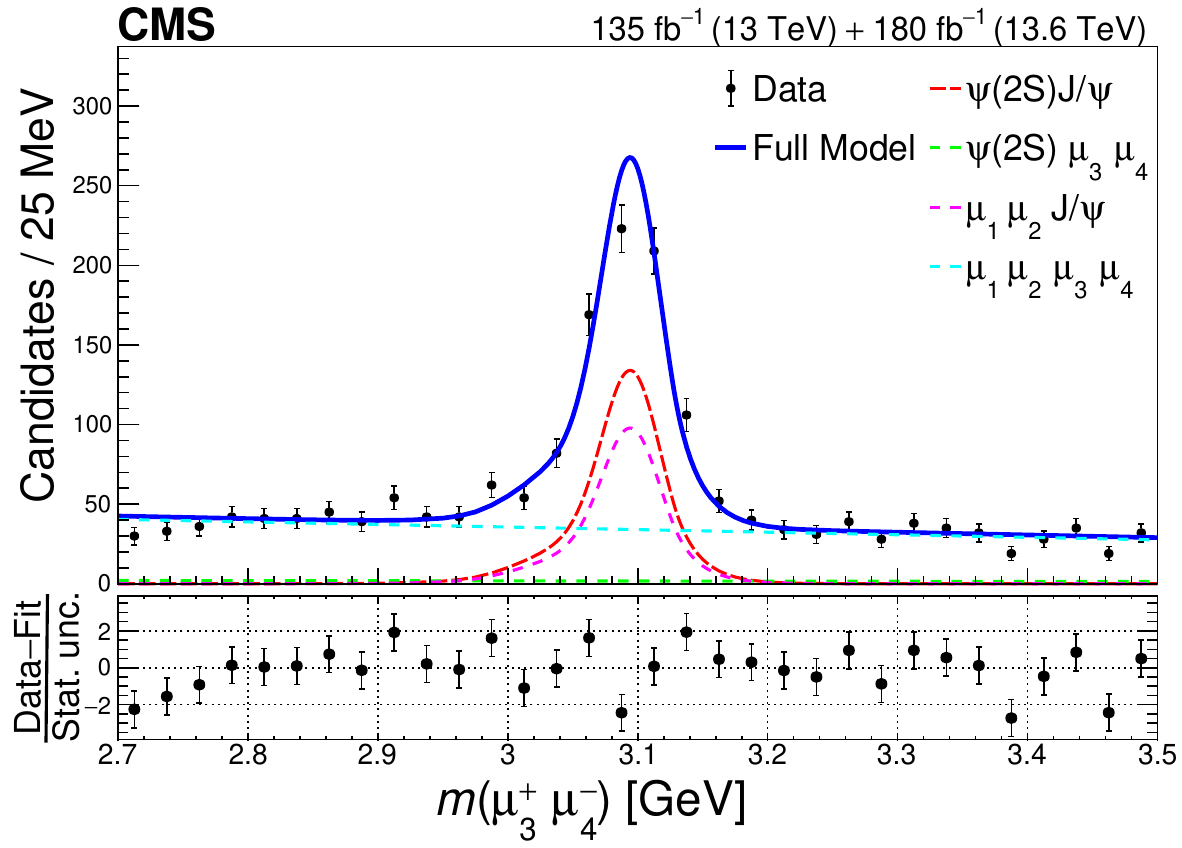}
\caption{
    \firstsentence{Projections of the 2D fit of \JPp (Fig.~\ref{3Dscattering})
    of selected pairs with $4\PGm$ masses below 15\GeV.} The \Jpsi and \Pps mass requirements and the mass constraints have been removed.
    The breakdown of the signal and different background components is shown.
    Left: Projection onto the \PGyP{2S} mass reconstructed from $\PGm_1^{+}\PGm_2^{-}$.
    Right: Projection onto the \Jpsi mass reconstructed from $\PGm_3^{+}\PGm_4^{-}$.
}
\label{2dfit_topup}
\end{figure*}

Because of the small size of the \JPp sample, no multiple candidates are observed per event. 
However, since the \JJ production rate is comparatively large, approximately 1\% of these events contaminate the \JPp selection via an incorrect muon pairing. 
We veto \JPp candidates that satisfy all 
criteria, but have an alternate pairing of muons where 
each pair falls within $2\sigma_m$ of the \Jpsi mass.  
A 2D distribution of the \PGyP{2S} versus \Jpsi candidate masses is shown in Fig.~\ref{3Dscattering}. The corresponding mass projections are shown in Fig.~\ref{2dfit_topup}, along with the 2D fit results, including the subcomponents of the fit.

\subsection{Event simulation}
Monte Carlo event generators are used to simulate events for a variety of purposes.
Signal MC samples are used to model the detector mass resolution and efficiency corrections.
The \textsc{JHUGen} generator (version~7.40)~\cite{1001.3396,1208.4018} is employed to simulate the process $\Pp\Pp\to \PX{+}\text{(anything)}$, where \PX denotes $J^P {\,=\,} 0^+,2^+$ mesons of different masses with negligible natural width.
These mesons subsequently decay into \JJ or \JPp with a $\PGmp\PGmm\PGmp\PGmm$ final state.
The \JJ or \JPp events from NRSPS, \ie, pairs arising from a single parton-parton collision without an intermediate resonance, are produced using the \PYTHIA generator (version 8.240)~\cite{Sjostrand:2014zea}. 
Feed-down processes from $\PX \to \JPp$, where the \Pps subsequently decays to \Jpsi with additional particles, are also simulated using the \PYTHIA generator, with \PX corresponding to \Xsn and \Xso.
The simulation incorporates the impact of additional $\Pp\Pp$ interactions that occur at the same time or close in time to the main collision.
Subsequently, the generated events are processed through the {CMS} detector simulation based on \GEANTfour~\cite{GEANT4:2002zbu}, followed by the standard reconstruction chain.

\subsection{Background models}
The background model for the \JJ channel includes a near-threshold excess, combinatorial backgrounds, and contributions from NRSPS, DPS, and feed-down from decays of heavier tetraquarks.
For the \JPp channel, which has a higher kinematic threshold, the background is composed only of NRSPS, DPS, and combinatorial contributions.

Nonresonant backgrounds, which arise without a resonant state, are present in both the \JJ and \JPp channels and include contributions from single parton-parton scattering, NRSPS, as well as double parton scattering, DPS, where two independent parton-parton interactions occur within the same $\Pp\Pp$ collision.
The parametrization of NRSPS is from the distribution of simulated events using \PYTHIA as described above, with the exponential slope parameter free in the mass spectrum fit.
The DPS shape is derived from data via event mixing, in which two meson candidates from separate selected events are combined to form a mixed event, and the resulting mass spectrum is used to model the shape of the DPS background.

Combinatorial backgrounds arise when muon candidates do not originate from a genuine \Jpsi or \Pps decay: \ie, due to random pairings, misidentified hadrons, or misreconstructed muons. 
In the Run~2 study, this background was negligible because the L1 trigger required three muons, ensuring that three of the four muons were high-quality; any residual contribution could be absorbed into the NRSPS and DPS components.
In Run~3, however, the introduction of the parking stream~\cite{CMS:2024zhe} allowed a trigger requiring only two muons at L1. While this increased the signal yield, it also enhanced the combinatorial background, necessitating its explicit treatment in the analysis.
The background shape is extracted from data using the ``nine-tile'' method~\cite{CMD-3:2019cvx}.
In this method, the 2D dimuon mass plane is divided into $3{\times}3$ regions (``tiles'') based on the two dimuon masses, with the central tile corresponding to the signal region defined by the \Jpsi or \Pps mass requirement. 
The eight surrounding tiles, which contain different mixtures of combinatorial components, are used to interpolate the background shape into the signal region.
The yield is estimated through the 2D fit to the $m(\Jpsi_1)$ versus $m(\Jpsi_2)$ distribution, or $m(\Pps)$ versus $m(\Jpsi)$ in the \JPp case, and incorporated into the full maximum-likelihood fit as a Gaussian-constrained component. 
The fitted combinatorial yield lies well within the imposed constraint.

Another background for the \JJ channel is feed-down from higher-mass states, such as $\PX \to \JPp$, where \Pps can decay to, for example, $\PJGy\,\PGpp\PGpm$. 
Such $\Jpsi{+}\text{(anything)}$ modes constitute $61.5 \pm 0.7\%$ of \Pps decays~\cite{PhysRevD.110.030001}.
This contribution was studied in the Run~2 analysis~\cite{cmscollaboration2023observation}, which indicated a nonzero feed-down yield from \Xsn in the \JPp channel, while other possible sources were consistent with zero.
Because of its then-speculative nature, feed-down was treated as a systematic uncertainty.
Now feed-down is incorporated into the baseline model because of new evidence: ATLAS reported a significant near-threshold excess in the \JPp spectrum~\cite{ATLAS:2023bft,ATLAS:J2S-2025}, and the \JPp analysis of the current work reveals structures consistent with \Xsn and \Xso decays. 
Feed-down contributions are modeled using the \PYTHIA generator, incorporating generated resonances at 6847\MeV with a width of 191\MeV corresponding to \Xsn, as well as at 7134\MeV with a width of 97\MeV for \Xso, both decaying via $\PGy(2S) \to \PJGy{+}\text{(anything)}$.
Based on the significances of the states found by ATLAS and CMS, the \Xsn feed-down is included in the baseline \JJ fit model, while the \Xso contribution is treated as a systematic uncertainty.

Aside from the above backgrounds, all three LHC experiments reported a near-threshold excess of \JJ candidates above background~\cite{cmscollaboration2023observation,LHCb:2020bwg,ATLAS:2023bft} with no clear interpretation.
Given the uncertainty of its origin, it is incorporated into the background model and described by an \emph{ad hoc} BW function, $\mathrm{BW}_0$, which doesn't interfere with any other component.

\subsection{Systematic uncertainties}
The systematic uncertainty in a parameter due to a given effect is quantified by the deviation of that parameter from its nominal value when the effect is varied.
The sources of systematic uncertainty for the \JJ analysis, following Ref.~\cite{cmscollaboration2023observation}, include: 
different BW shapes (P- and D-wave, and alternative values for
the Blatt--Weisskopf factor~\cite{Hippel:1972, PhysRevD.48.1225, Blatt:1952ije}),
the NRSPS, DPS, and combinatorial background shapes,
detector mass resolution, efficiency corrections, 
feed-down contribution from \Xso,
and effects of removing the $\mathrm{BW_0}$ component from the fit.
The principal systematic effects for the \JJ results are summarized in Table~\ref{Tab:JJSys}. 
The total uncertainties are the sums of the individual contributions in quadrature.

\begin{table}[htb!]
    \topcaption{
    \firstsentence{The principal contributions to the systematic uncertainties for \Xss, \Xsn, and \Xso from the \JJ interference fit.} The values are in \MeVns.
    }
    \label{Tab:JJSys}
    \centering
    \renewcommand\arraystretch{1.2}
    \begin{tabular}{lcccccc} 
& \multicolumn{2}{c}{\Xss} & \multicolumn{2}{c}{\Xsn} & \multicolumn{2}{c}{\Xso} \\[-2pt]
Source & $\Delta m$ &  $\Delta \Gamma$ & $\Delta m$ &  $\Delta \Gamma$ & $\Delta m$ &  $\Delta \Gamma$ \\
        \hline
        Signal shape & 25 & 52 & 2 & 11 & 3 & 5  \\
        NRSPS shape &  3 & 7  & $<$1  & 1 & $<$1 & 5  \\
        DPS shape &  $<$1 &  5  & $<$1 &  $<$1 &  $<$1 & 1 \\
        Comb. bkg. shape & $<$1 & 22 & $<$1 & 2 & $<$1 & 4  \\
        Mass resolution & 4 & 58 & 15 & 7 & 12 & 5   \\
        Efficiency & $<$1 & 4  &$<$1  &$<$1  &$<$1  &$<$1   \\
        Feed-down & $<$1 &  1 & $<$1  &$<$1  & $<$1 & $<$1 \\
        Without $\mathrm{BW}_0$ & $<$1 & 29  & 2 & 3 & 2 & 1  \\[2pt]
        Total uncertainty & 25 & 87 & 15 & 14 & 13 & 10  \\
    \end{tabular}
\end{table}

A similar procedure is applied to estimate the \Xsn and \Xso systematic uncertainties for the \JPp analysis.
Based on the sources considered for the \JJ analysis, the $\mathrm{BW}_0$ and feed-down uncertainties are absent, while two additional systematic uncertainties are included: the potential impact of a \Xss tail and bias from the fit due to the small number of \JPp events.

Given the relatively large width of the \Xss observed in the \JJ spectrum, its tail extends into the \JPp region.
To account for this, an \Xss BW function is added to the \JPp model with a three-peak interference fit, where the \Xss mass, width, and amplitude are fixed to the values determined in the \JJ analysis, while the relative phase is allowed to float. 
Parameter shifts of this fit relative to the \JPp fit without the \Xss component provide the systematic uncertainties due to \Xss effects.

Fit bias in the \JPp channel, arising from the small event sample, is evaluated using toy MC studies generated from the nominal fit parameters. 
Pull distributions, defined as the difference between fitted and nominal values divided by the fitted uncertainty, are analyzed. 
These distributions are often asymmetric and do not always follow a standard normal distribution. 
To account for this, each pull distribution is fitted with a sum of two asymmetric (bifurcated) Gaussian functions that share a common mean, from which we extract the mean as well as the left- and right-side widths.
The pull fit results are then converted into asymmetric systematic uncertainties. 
The effective right-side sigma is calculated as a weighted quadrature average of the right-side sigmas of the two asymmetric Gaussian components, with the weights given by the fraction of each component in the fit.
The positive systematic uncertainty is then obtained by scaling the statistical uncertainty with the sum of the pull mean and the effective right-side sigma, and subtracting the original statistical uncertainty. 
A similar procedure, using the effective left-side sigma, is applied to derive the negative systematic uncertainty.

The principal \JPp systematic effects are summarized in Table~\ref{Tab:Sys}.
The different characteristics of the \JJ and \JPp spectra ($\eg$, statistical power, proximity to threshold, etc.) result in some systematic sources having different degrees of importance. In particular, the DPS shape systematic uncertainty is negligible for the \JPp fit and is therefore not included in Table~\ref{Tab:Sys}. 

\begin{table}[htb!]
\topcaption
{
    \firstsentence{The principal contributions to the systematic uncertainties for \Xsn and \Xso from the \JPp interference fit.} The values are in \MeVns.
}
\label{Tab:Sys}
\centering
\renewcommand\arraystretch{1.2}
    \begin{tabular}{lcccc}
 & \multicolumn{2}{c}{\Xsn} & \multicolumn{2}{c}{\Xso} \\[-2pt]
    Source & $\Delta m$ & $\Delta \Gamma$  & $\Delta m$  & $\Delta \Gamma$  \\
    \hline
    Signal shape & 29 & 79 & 22 & 131\\
    NRSPS shape & 14 & 54 & 14 & 29 \\
    Comb. bkg. shape & 15& 51& 15 & 20 \\
    Mass resolution & 5 & 7 & 5 & 9\\
    Efficiency & 7 & 27 & 7 & 10 \\
    Fitter bias & $^{+16}_{-17}$  &  $^{+83}_{-72}$ & $^{+41}_{-22}$ & $^{+3}_{-115}$ \\
    \Xss addition & 104 & 14 & 61 & 31 \\[2pt]
    \multirow{2}{*}{Total uncertainty} & $+111$ & $+140$ & $+80$ & $+140$ \\[-4pt]
     & $-112$ & $-134$ & $-72$ & $-181$ \\
    \end{tabular}
\end{table}

\begin{acknowledgments}
We congratulate our colleagues in the CERN accelerator departments for the excellent performance of the LHC and thank the technical and administrative staffs at CERN and at other CMS institutes for their contributions to the success of the CMS effort. In addition, we gratefully acknowledge the computing centers and personnel of the Worldwide LHC Computing Grid and other centers for delivering so effectively the computing infrastructure essential to our analyses. Finally, we acknowledge the enduring support for the construction and operation of the LHC, the CMS detector, and the supporting computing infrastructure provided by the following funding agencies: SC (Armenia), BMBWF and FWF (Austria); FNRS and FWO (Belgium); CNPq, CAPES, FAPERJ, FAPERGS, and FAPESP (Brazil); MES and BNSF (Bulgaria); CERN; CAS, MoST, and NSFC (China); MINCIENCIAS (Colombia); MSES and CSF (Croatia); RIF (Cyprus); SENESCYT (Ecuador); ERC PRG, TARISTU24-TK10 and MoER TK202 (Estonia); Academy of Finland, MEC, and HIP (Finland); CEA and CNRS/IN2P3 (France); SRNSF (Georgia); BMFTR, DFG, and HGF (Germany); GSRI (Greece); NKFIH (Hungary); DAE and DST (India); IPM (Iran); SFI (Ireland); INFN (Italy); MSIT and NRF (Republic of Korea); MES (Latvia); LMTLT (Lithuania); MOE and UM (Malaysia); BUAP, CINVESTAV, CONACYT, LNS, SEP, and UASLP-FAI (Mexico); MOS (Montenegro); MBIE (New Zealand); PAEC (Pakistan); MES, NSC, and NAWA (Poland); FCT (Portugal); MESTD (Serbia); MICIU/AEI and PCTI (Spain); MOSTR (Sri Lanka); Swiss Funding Agencies (Switzerland); MST (Taipei); MHESI (Thailand); TUBITAK and TENMAK (T\"{u}rkiye); NASU (Ukraine); STFC (United Kingdom); DOE and NSF (USA).

\hyphenation{Rachada-pisek} Individuals have received support from the Marie-Curie program and the European Research Council and Horizon 2020 Grant, contract Nos.\ 675440, 724704, 752730, 758316, 765710, 824093, 101115353, 101002207, 101001205, and COST Action CA16108 (European Union); the Leventis Foundation; the Alfred P.\ Sloan Foundation; the Alexander von Humboldt Foundation; the Science Committee, project no. 22rl-037 (Armenia); the Fonds pour la Formation \`a la Recherche dans l'Industrie et dans l'Agriculture (FRIA) and Fonds voor Wetenschappelijk Onderzoek contract No. 1228724N (Belgium); the Beijing Municipal Science \& Technology Commission, No. Z191100007219010, the Fundamental Research Funds for the Central Universities, the Ministry of Science and Technology of China under Grant No. 2023YFA1605804, the Natural Science Foundation of China under Grant No. 12535004, and USTC Research Funds of the Double First-Class Initiative No.\ YD2030002017 (China); the Ministry of Education, Youth and Sports (MEYS) of the Czech Republic; the Shota Rustaveli National Science Foundation, grant FR-22-985 (Georgia); the Deutsche Forschungsgemeinschaft (DFG), among others, under Germany's Excellence Strategy -- EXC 2121 ``Quantum Universe" -- 390833306, and under project number 400140256 - GRK2497; the Hellenic Foundation for Research and Innovation (HFRI), Project Number 2288 (Greece); the Hungarian Academy of Sciences, the New National Excellence Program - \'UNKP, the NKFIH research grants K 131991, K 133046, K 138136, K 143460, K 143477, K 146913, K 146914, K 147048, 2020-2.2.1-ED-2021-00181, TKP2021-NKTA-64, and 2025-1.1.5-NEMZ\_KI-2025-00004 (Hungary); the Council of Science and Industrial Research, India; ICSC -- National Research Center for High Performance Computing, Big Data and Quantum Computing, FAIR -- Future Artificial Intelligence Research, and CUP I53D23001070006 (Mission 4 Component 1), funded by the NextGenerationEU program (Italy); the Latvian Council of Science; the Ministry of Education and Science, project no. 2022/WK/14, and the National Science Center, contracts Opus 2021/41/B/ST2/01369, 2021/43/B/ST2/01552, 2023/49/B/ST2/03273, and the NAWA contract BPN/PPO/2021/1/00011 (Poland); the Funda\c{c}\~ao para a Ci\^encia e a Tecnologia, grant CEECIND/01334/2018 (Portugal); the National Priorities Research Program by Qatar National Research Fund; MICIU/AEI/10.13039/501100011033, ERDF/EU, ``European Union NextGenerationEU/PRTR", and Programa Severo Ochoa del Principado de Asturias (Spain); the Chulalongkorn Academic into Its 2nd Century Project Advancement Project, the National Science, Research and Innovation Fund program IND\_FF\_68\_369\_2300\_097, and the Program Management Unit for Human Resources \& Institutional Development, Research and Innovation, grant B39G680009 (Thailand); the Kavli Foundation; the Nvidia Corporation; the SuperMicro Corporation; the Welch Foundation, contract C-1845; and the Weston Havens Foundation (USA).
\end{acknowledgments}\section*{Data availability} Release and preservation of data used by the CMS Collaboration as the basis for publications is guided by the  \href{https://doi.org/10.7483/OPENDATA.CMS.1BNU.8V1W}{CMS data preservation, re-use and open access policy}.
\bibliography{auto_generated}

\ifthenelse{\boolean{cms@external}}{}
{
\appendix
\numberwithin{figure}{section}
\numberwithin{table}{section}
\section{Supplementary Materials\label{app:Methods}}

\input{supplemental_material.tex}

}

\cleardoublepage \section{The CMS Collaboration \label{app:collab}}\begin{sloppypar}\hyphenpenalty=5000\widowpenalty=500\clubpenalty=5000\input{BPH-24-003-public-authorlist.tex}\end{sloppypar}
\end{document}

%% file: supplemental_material.tex
\subsection{The full-range mass spectra fits}
\begin{figure*}[htbp]
\centering
\includegraphics[width=0.6\textwidth]{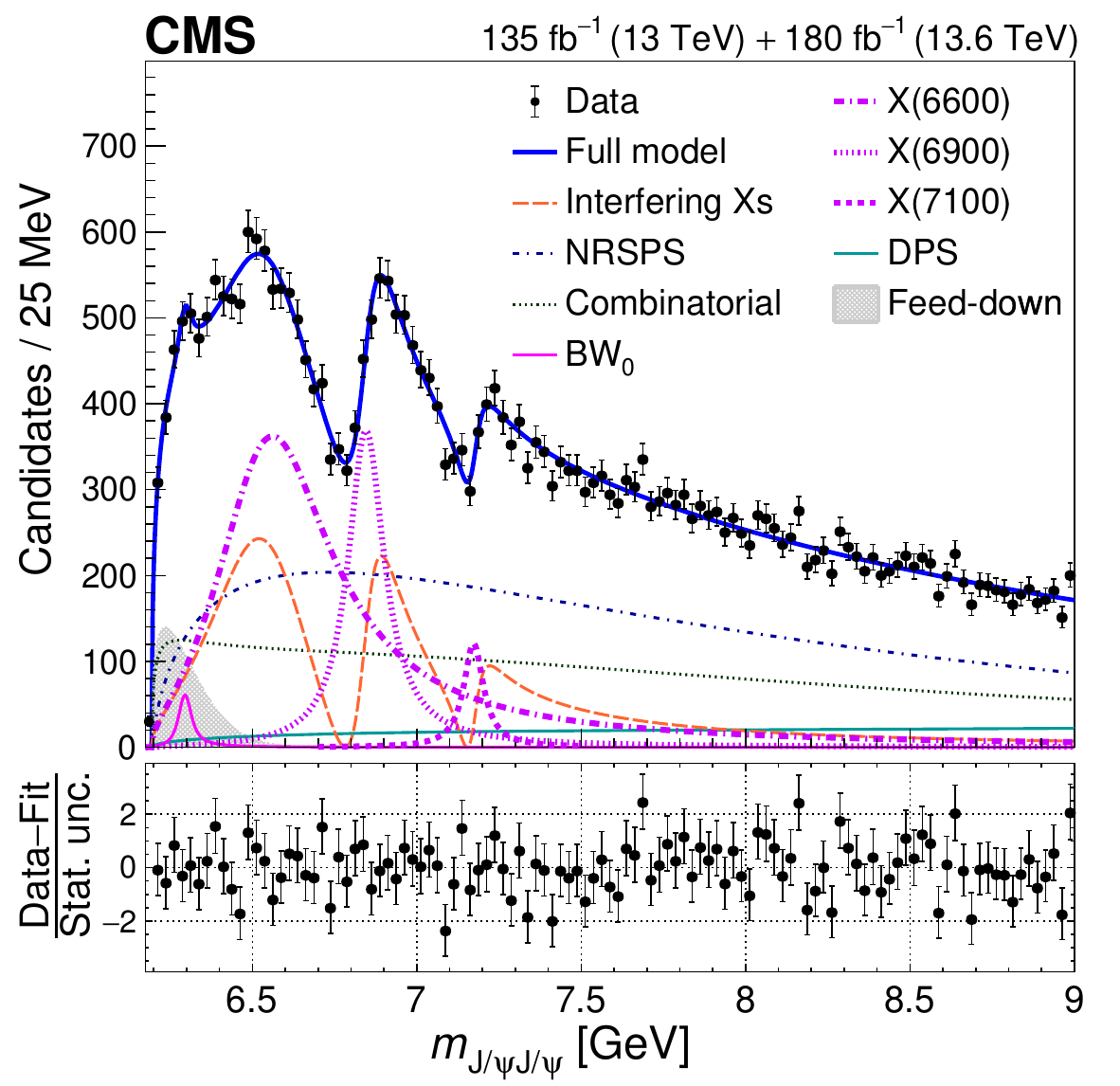}
\includegraphics[width=0.99\textwidth]{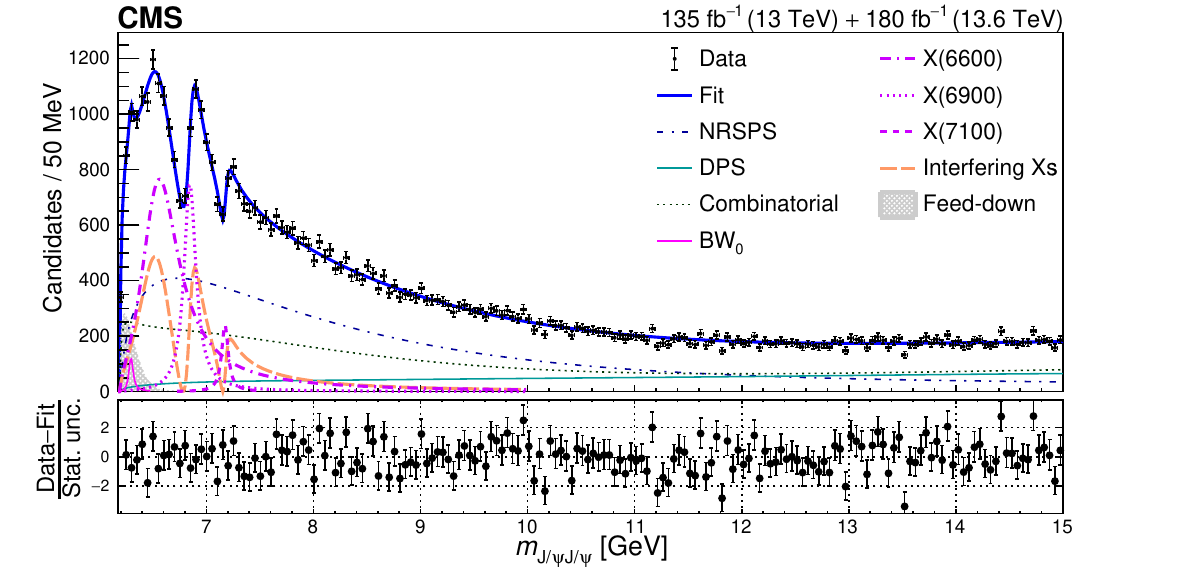}
\caption
{
    \firstsentence{\JJ invariant-mass spectrum for the interference fit.}
    The spectrum is shown up to $9\GeV$ (top) and over the full fit range (bottom).
    Three interfering BW signals are included, and the total contribution from all interfering amplitudes (including cross terms) is plotted as the curve labeled ``Interfering Xs''. 
    The background components---NRSPS, DPS, combinatorial, feed-down, and threshold $\mathrm{BW}_0$---are shown separately.
    The lower panels give the number of standard deviations (statistical uncertainties only) that the binned data deviates from the fits.
}
\label{Fig:full_range}
\end{figure*}

\begin{figure*}[htbp]
\centering
\includegraphics[width=0.6\textwidth]{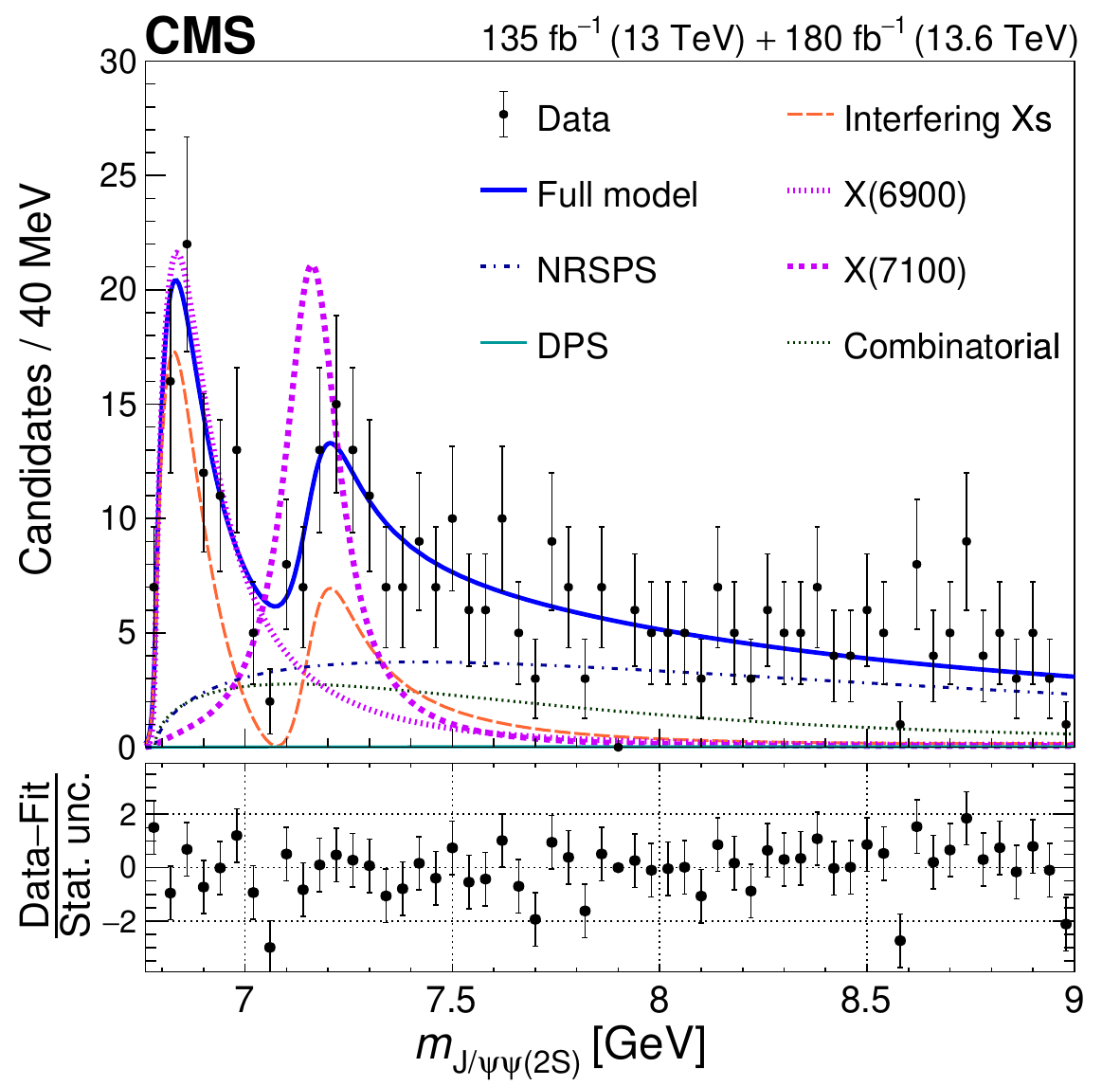}
\includegraphics[width=0.99\textwidth]{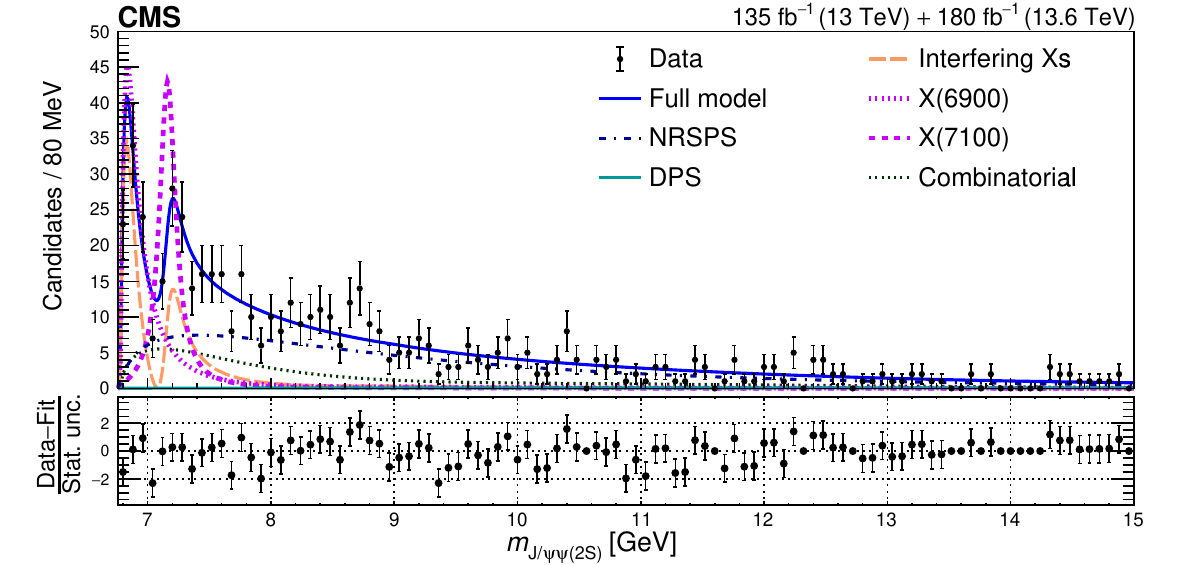}
\caption
{
    \firstsentence{\JPp invariant-mass spectrum for the interference fit.}
    The spectrum is shown up to $9\GeV$ (top) and over the full fit range (bottom).
    Three interfering BW signals are included, and the total contribution from all interfering amplitudes (including cross terms) is plotted as the curve labeled ``Interfering Xs''. 
    The background components are shown separately.
    The lower panel also displays the number of standard deviations (statistical uncertainties only) that the binned data deviates from the fits.
}
\label{Fig:JP_full_range}
\end{figure*}

The fits to the \JJ and \JPp mass spectra are performed from threshold up to 15\GeV.  
To better identify the \PX structures and interference, Fig.~\insupp{3 in the main document}{Fig:Interf-fitmodel} only extends to 9\GeV and the background portion of the fit is shown as a sum of the components.  
Figure~\ref{Fig:full_range} shows the \JJ interference fit with individual background components, displayed up to 9\GeV (top) using the same binning as Fig.~\insupp{3 in the main document}{Fig:Interf-fitmodel}, and over the full mass range (bottom) with a larger bin size for clarity.
The fitted parameters are the masses and widths of the four BW functions, the relative amplitudes and phases of the \Xss and \Xso contributions, a shape parameter for the NRSPS component, the yields of the five background contributions, and the yield of the three interfering signals.
The fit has 19 free parameters and the resulting quality is characterized by a $\chi^2$ per degree of freedom of 346/334 over the full mass range (Fig.~\ref{Fig:full_range}), calculated using the same binning as in Fig.~\insupp{3 in the main document}{Fig:Interf-fitmodel}. 
Because the signal region constitutes only a small fraction of the full spectrum, localized deviations are diluted by the broad high-mass continuum.
To better assess the signal region, we compute the $\chi^2/\nbin$ below 7.8\GeV, obtaining 52/65, with \nbin as the number of nonzero bins in the restricted range.
Similar plots for the \JPp events are provided in Fig.~\ref{Fig:JP_full_range}.

\subsection{Alternate noninterference \texorpdfstring{\JJ}{J/psi J/psi} mass spectrum fit}
\begin{figure}[htbp]
    \centering
    \includegraphics[width=0.49\textwidth]{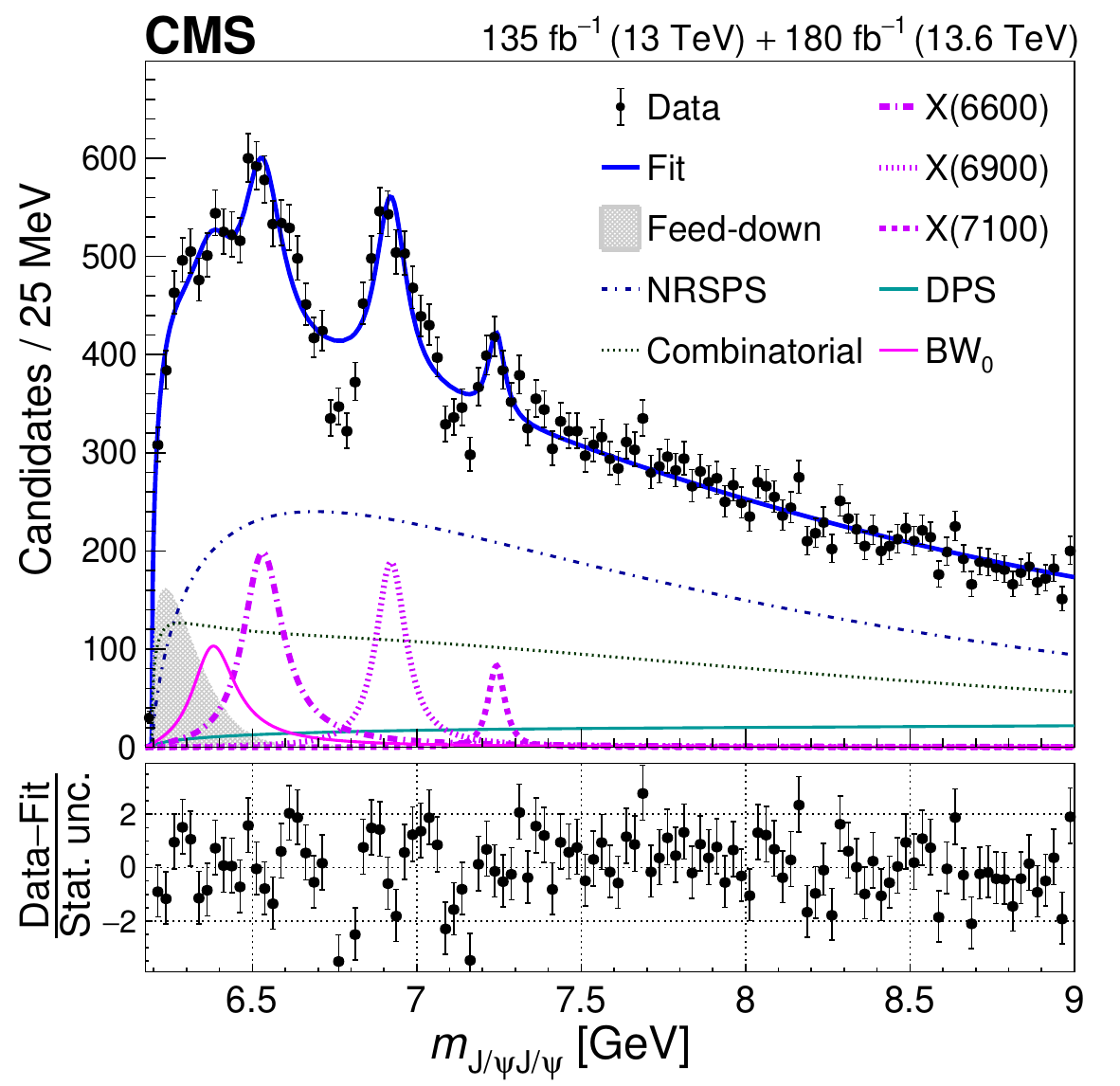}
    \caption{\firstsentence{The noninterference fit result of the \JJ invariant-mass spectrum up to 9\GeV.}
    The lower panel shows the deviation of the data from the fit, in units of standard deviations.
    }
\label{Fig:NonInterf-fitmodel}
\end{figure}

A fit with no interference, following Ref.~\cite{cmscollaboration2023observation}, was also tested. 
In this model, none of the components interfere; each resonance is described by $\abs{\mathrm{BW}}^2 \otimes R(M)$, corresponding to a purely noninterfering configuration.
The noninterference fit model provides a poor description of the ``dips'' around 6750 and 7150\MeV, as shown in Fig.~\ref{Fig:NonInterf-fitmodel}.
The numerical results of the noninterference fit of the \JJ mass spectrum are given in Table~\ref{Tab:fulltable}, along with the interference fit. 
The BW masses significantly shift, reflecting the fact that destructive interference shifts the visible peak maxima off the resonance masses.
The noninterference widths are narrower, which is driven by the minimization trying to get the BW functions to follow the dips.
For the noninterference fit, the $\chi^2/\nbin$ below 7.8\GeV is 147/65.

\begin{table}[htbp]
    \topcaption{\firstsentence{Fit results of the Run~2+3 \JJ mass spectra with the baseline interference model (reproduced from main text), and a noninterference model.}
    Masses and widths are in \MeVns (double uncertainties are statistical followed by systematic; single uncertainties are statistical only).}
  \centering
    \renewcommand\arraystretch{1.2} 
  \begin{tabular}{c c c c c c}
        Resonance & Value & Interference & Noninterference \\
        \hline
            \multirow{2}{*}{\Xss}
            &$m$      & $6593_{\,-\,14}^{\,+\,15} \pm 25$ & $6540 \pm 8$ \\
            &$\Gamma$ & $446_{\,-\,54}^{\,+\,66} \pm 87$  & $142 \pm 19$ \\[4pt]
            \multirow{2}{*}{\Xsn}
            &$m$      & $6847 \pm 10 \pm 15$      & $6927 \pm 6$ \\
            &$\Gamma$ & $135_{\,-\,14}^{\,+\,16} \pm 14$  & $97 \pm 13$  \\[4pt]
            \multirow{2}{*}{\Xso}
            &$m$      & $7173_{\,-\,10}^{\,+\,9} \pm 13$  & $7247 \pm 8$ \\
            &$\Gamma$ & $73_{\,-\,15}^{\,+\,18} \pm 10$   & $37 \pm 24$  \\
\end{tabular}
    \label{Tab:fulltable}
\end{table}

\subsection{Comparison of the \texorpdfstring{\PX}{X} family with other Regge predictions}
Many $J^{PC}$ quantum numbers are in principle possible for the \PX states, but the observed decay of $\PX \to \JJ$ restricts $C$ to be positive. 
The spin quantum number $J$ of the \PX is determined by the combination of the spins $S$ and orbital angular momentum $L$ of the constituent particles.
It is convenient to organize the quantum numbers of the four quarks in terms of quark pairs, either $\PQc\PQc/\PAQc\PAQc$ pairs in diquark models, or $\PQc\PAQc$ pairs in molecular models. 
Because the theoretical calculations that follow are all diquark models, we restrict our discussion to that case, but quantum numbers for molecular models follow the same patterns.
The simplest configurations of quark pairs have zero orbital angular momentum, and quark spins are either anti-aligned or aligned for the respective diquark spins of $0$ or $1$. 
Potential \PX states are composed of either spin $0$ or $1$ diquarks, with relative angular momentum $L= 0,$  1,  2, or higher, and total diquark spin $S= 0$,  1,  2.
The internal configuration of an $\PX$ state is thus specified by $J$, $L$, and $S$.

\begin{figure*}[htbp]
\centering
\includegraphics[width=0.45\textwidth]{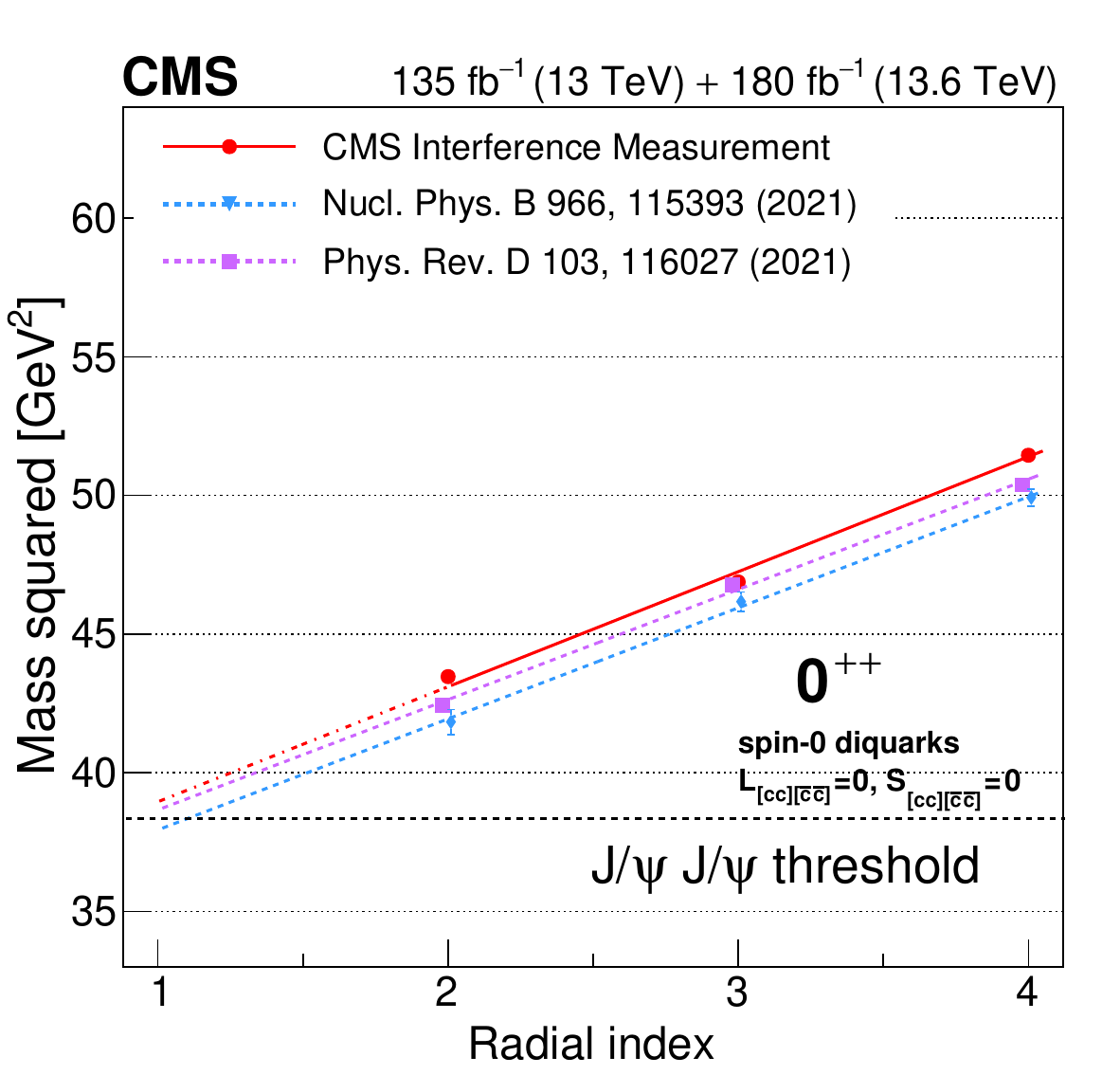}
\includegraphics[width=0.45\textwidth]{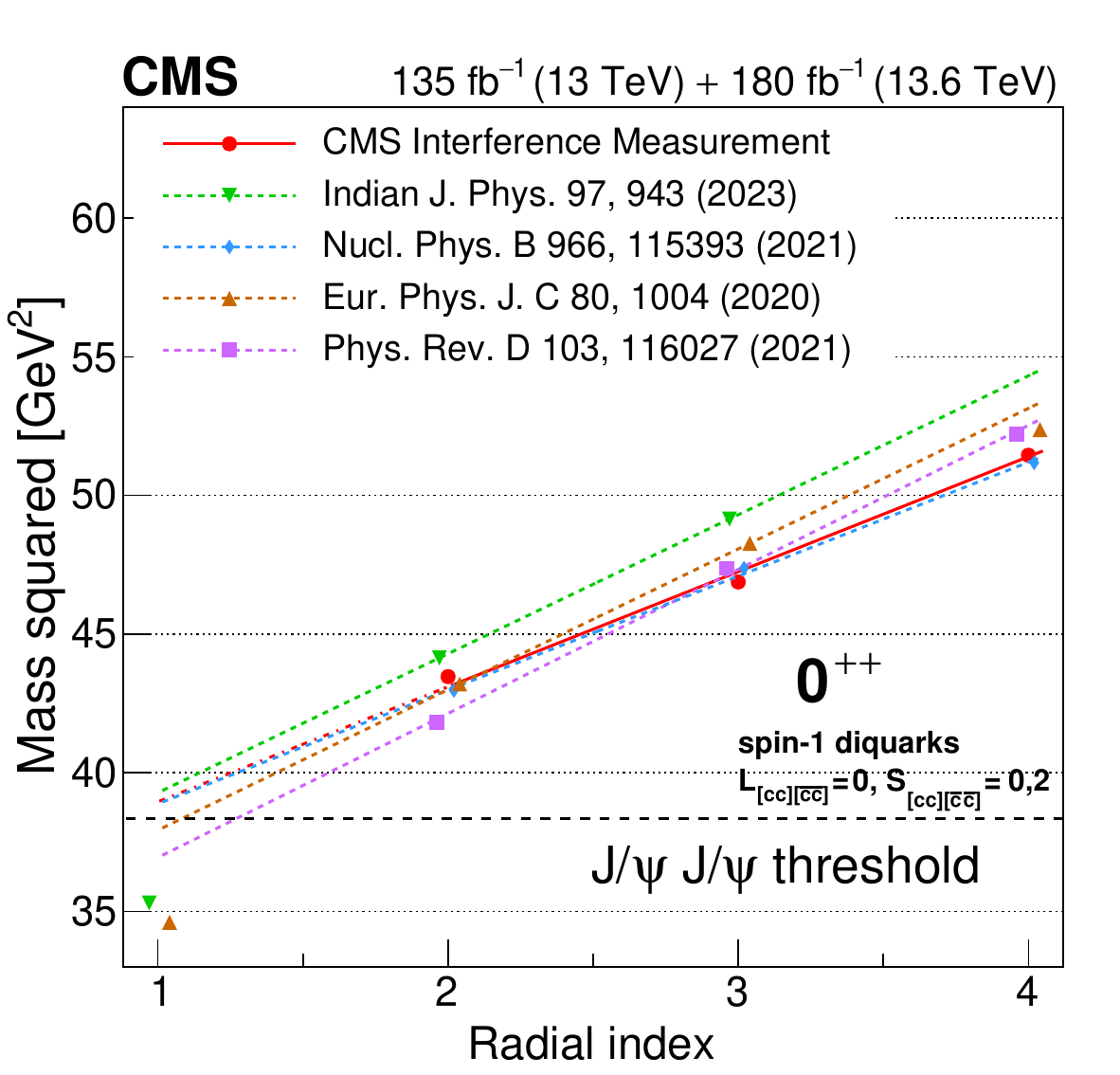}
\includegraphics[width=0.45\textwidth]{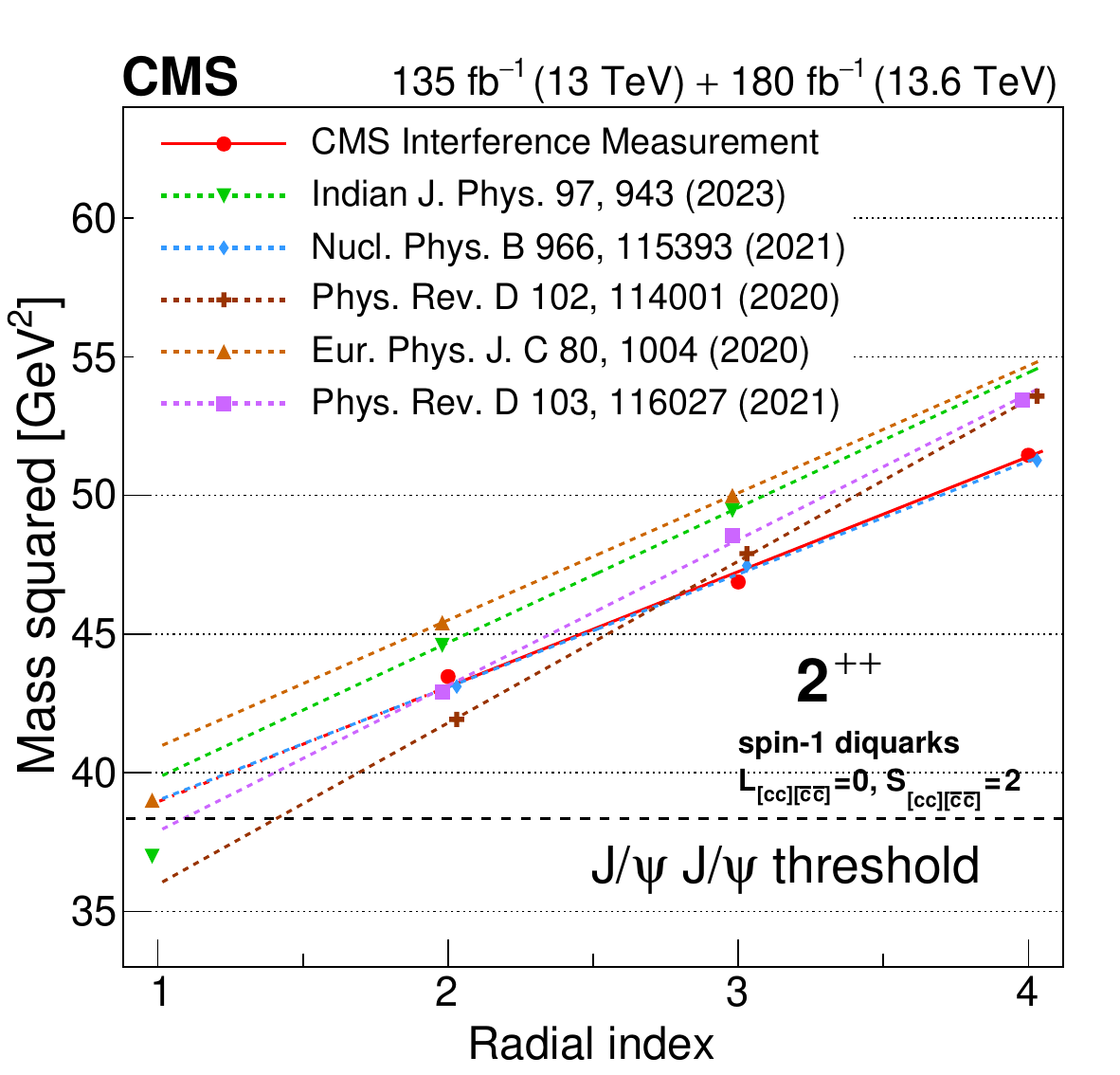}
\caption
{
    \firstsentence{Comparison of the measured \PX Regge trajectory to theoretical calculations of trajectories for various $J^{PC}$ quantum numbers corresponding to $L=0$ states~\cite{Zhu:2020xni,Zhao:2020jvl,Tiwari:2021tmz,Bedolla:2019zwg}.}
    Dashed lines show Regge trajectories fitted to the calculated squared masses, with the \PX trajectory (solid line) overlaid for comparison (for visual clarity, no uncertainty bands are shown).
    The spin-0 and spin-1 labels correspond to scalar and axial-vector diquark configurations, respectively.
    $L_{[\PQc\PQc][\PAQc\PAQc]}$ is the orbital angular momentum between the diquark and antidiquark, and $S_{[\PQc\PQc][\PAQc\PAQc]}$ is the total spin of the diquark-antidiquark system.
}
\label{fig:libraryRegge_Swave}
\end{figure*}

\begin{figure*}[htbp]
\centering
\includegraphics[width=0.45\textwidth]{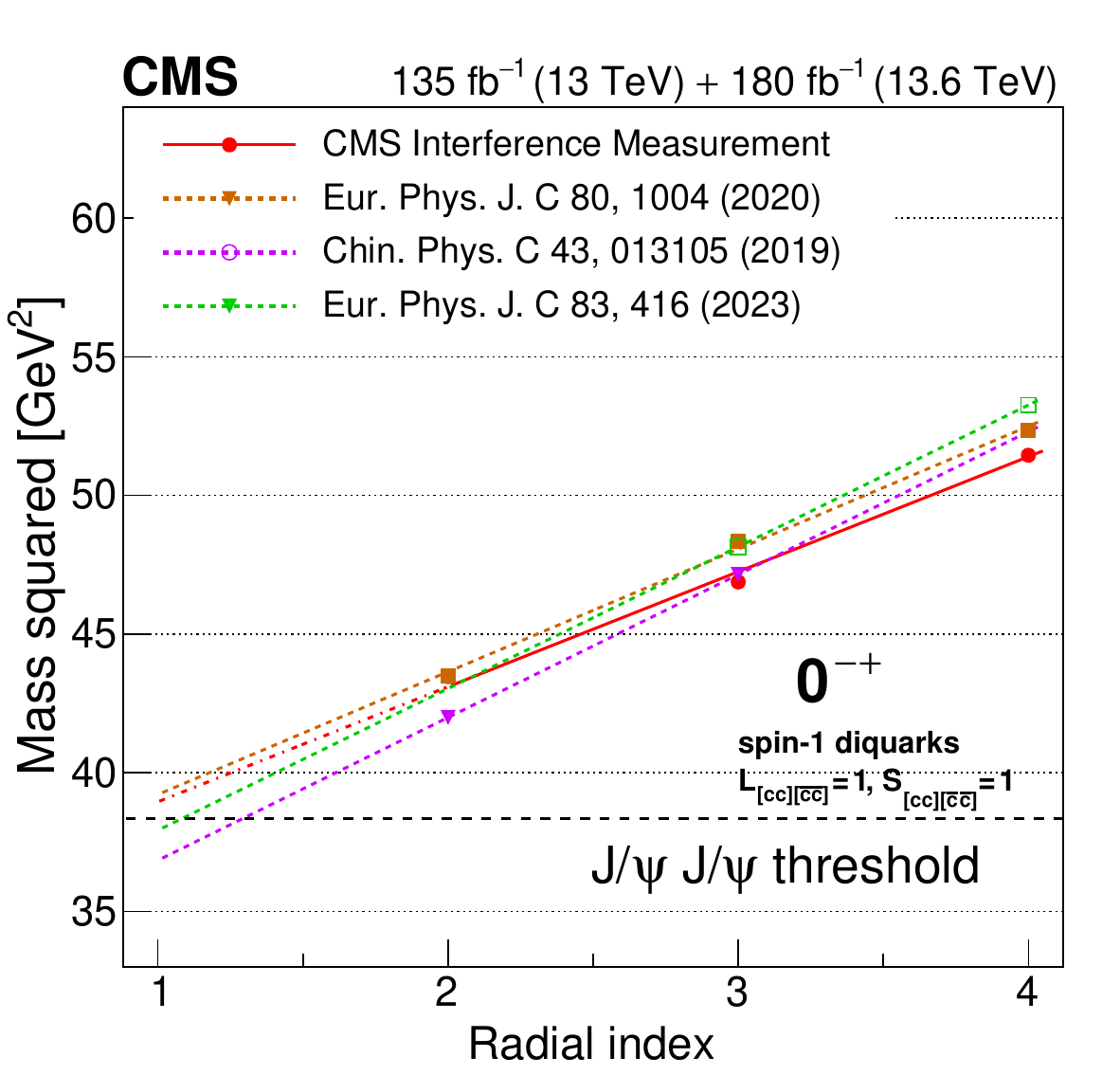}
\includegraphics[width=0.45\textwidth]{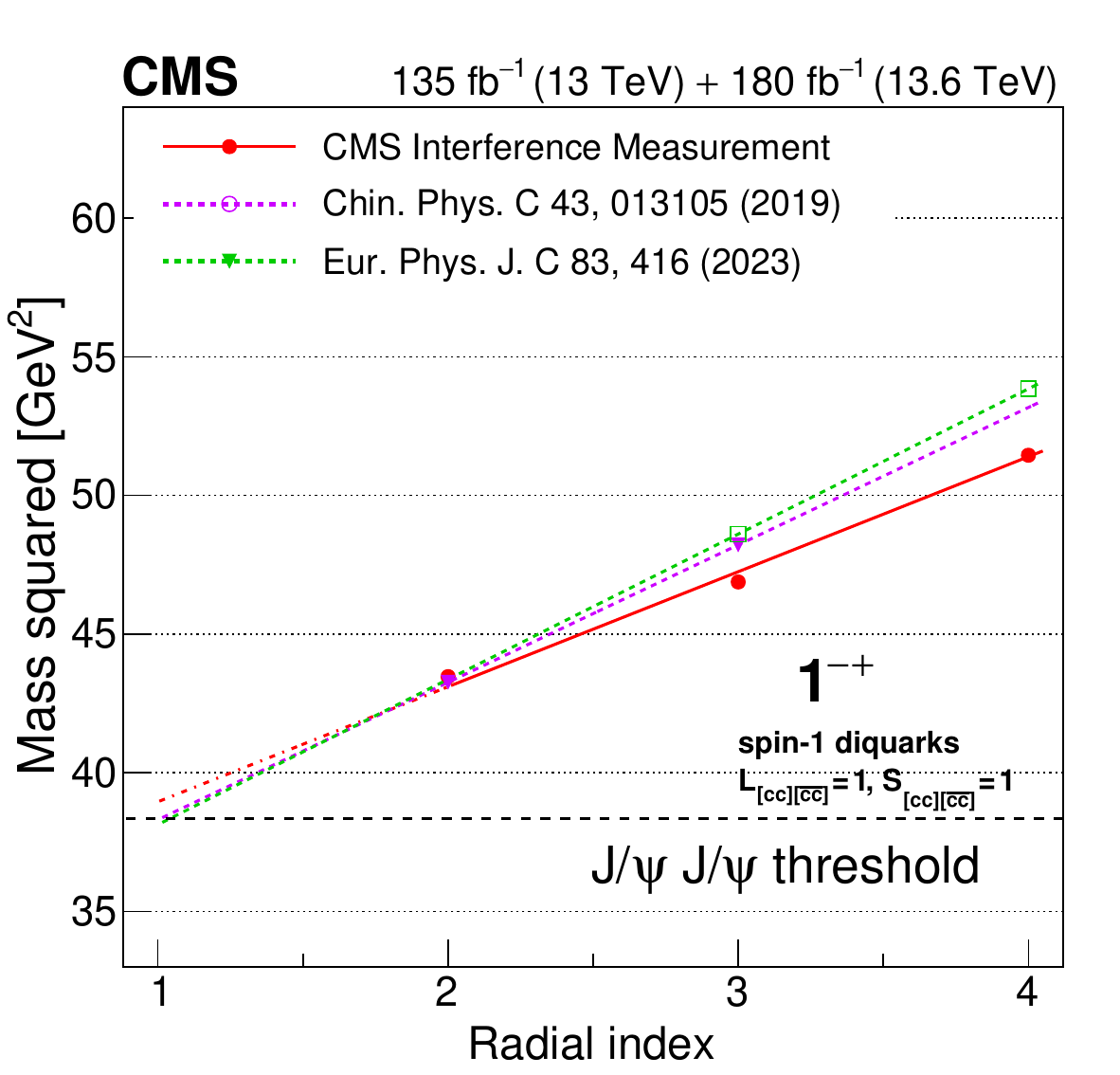}
\includegraphics[width=0.45\textwidth]{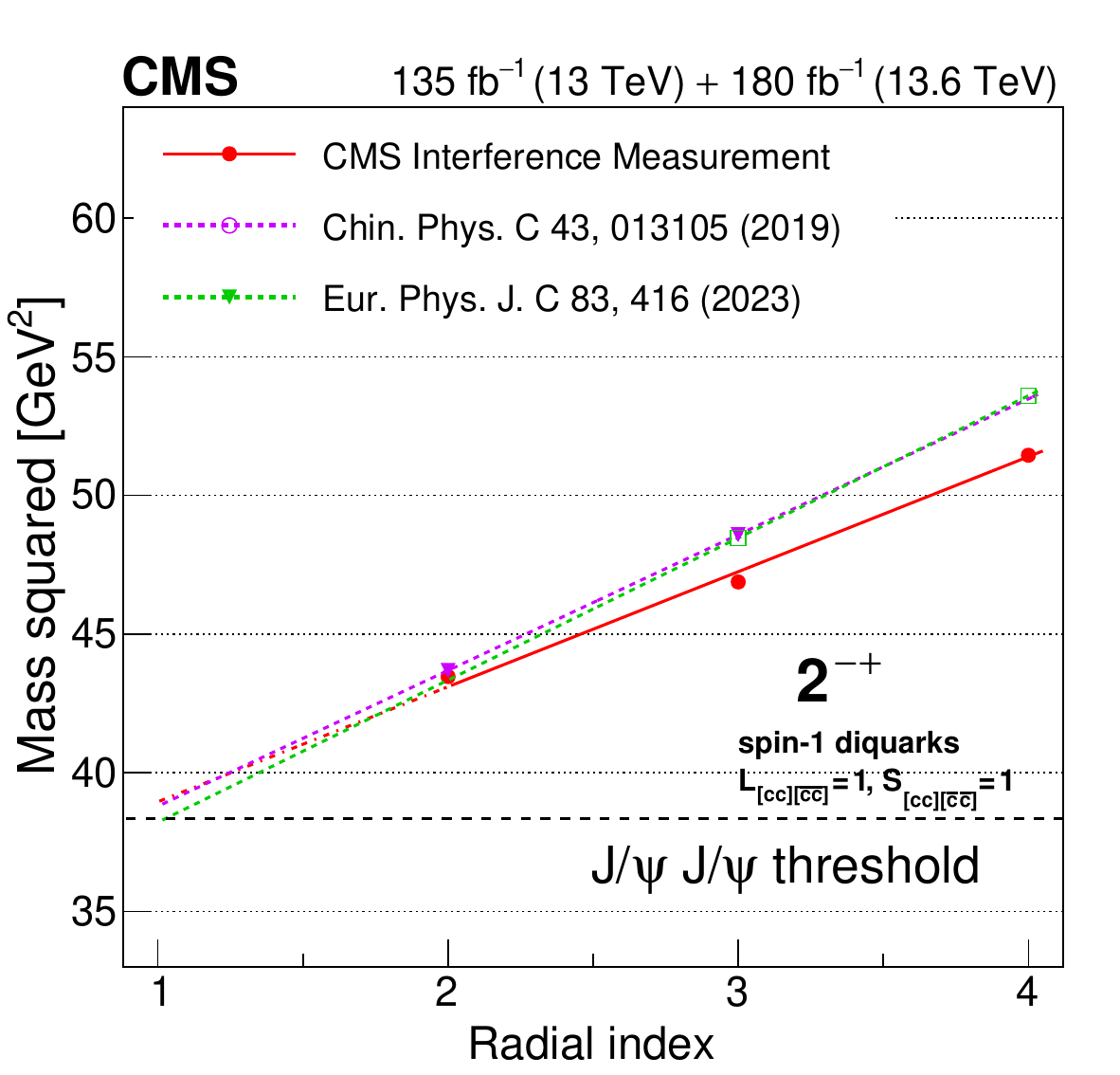}
\caption
{
    \firstsentence{Comparison of the measured \PX Regge trajectory to theoretical calculations of trajectories for various $J^{PC}$ quantum numbers corresponding to $L=1$ states~\cite{Bedolla:2019zwg,Yu:2022lak,Debastiani:DiquarkSize2017}.}
    Dashed lines show Regge trajectories fitted to the calculated squared masses, with the \PX trajectory (solid line) overlaid for comparison (for visual clarity, no uncertainty bands are shown).
    The diquarks are in an axial-vector configuration.
    $L_{[\PQc\PQc][\PAQc\PAQc]}$ is the orbital angular momentum between the diquark and antidiquark, and $S_{[\PQc\PQc][\PAQc\PAQc]}$ is the total spin of the diquark-antidiquark system.
}
\label{fig:libraryRegge_Pwave}
\end{figure*}

\begin{figure}[htbp]
\centering
\includegraphics[width=0.45\textwidth]{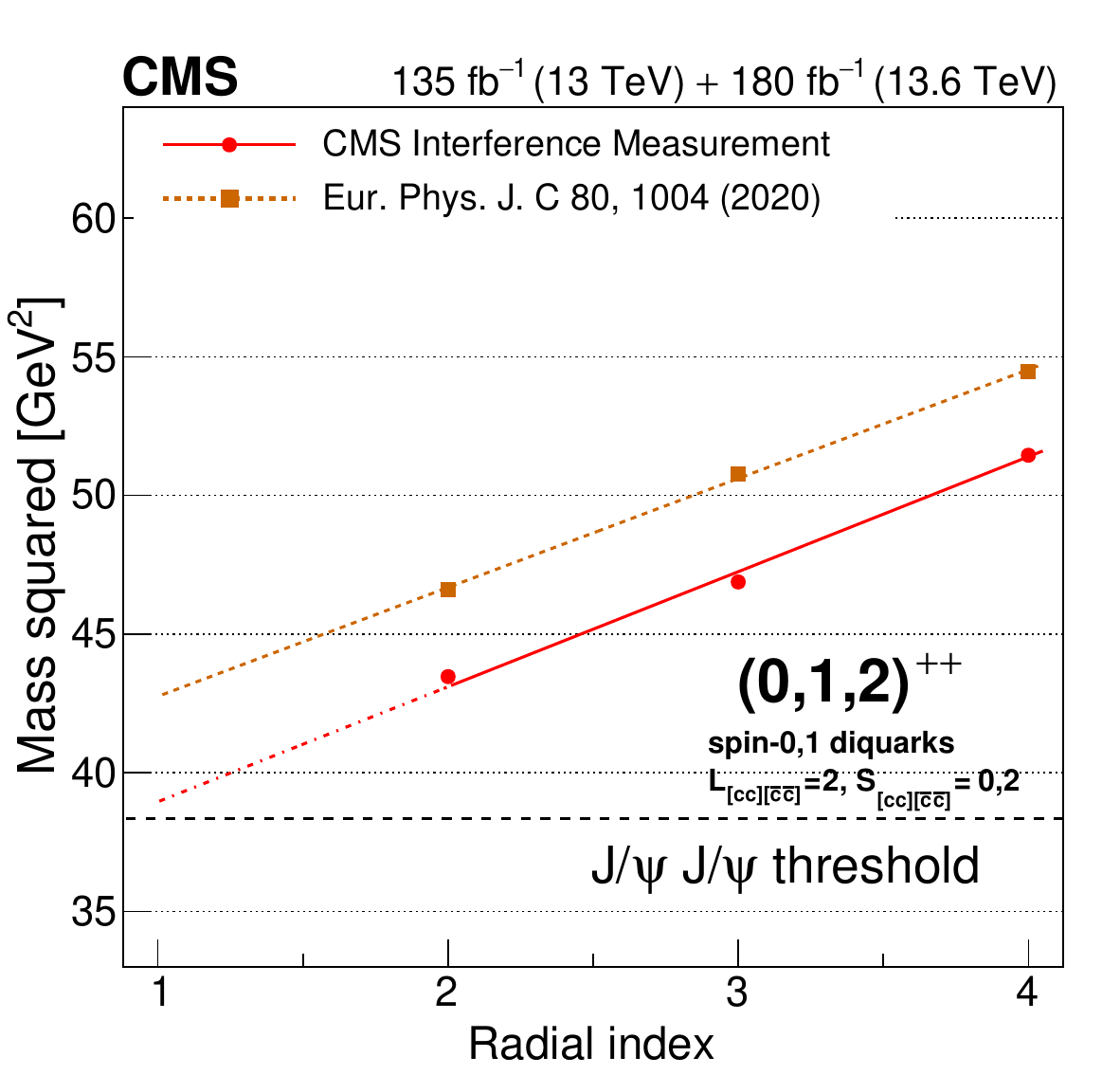}
\caption
{
    \firstsentence{Comparison of the measured \PX Regge trajectory to theoretical calculations of trajectories for various $J^{PC}$ quantum numbers corresponding to $L=2$ states~\cite{Bedolla:2019zwg}.}
    Dashed lines show Regge trajectories fitted to the calculated squared masses, with the \PX trajectory (solid line) overlaid for comparison (for visual clarity, no uncertainty bands are shown).
    The spin-0 and spin-1 labels correspond to scalar and axial-vector diquark configurations, respectively.
    $L_{[\PQc\PQc][\PAQc\PAQc]}$ is the orbital angular momentum between the diquark and antidiquark, and $S_{[\PQc\PQc][\PAQc\PAQc]}$ is the total spin of the diquark-antidiquark system.
}
\label{fig:libraryRegge_Dwave}
\end{figure}

The $0^{++}$ and $2^{++}$ $J^{PC}$ assignments are popular choices in the literature for the \PX's, but there are others available. 
We have collected diquark calculations of tetraquark masses~\cite{Zhu:2020xni, Tiwari:2021tmz, Zhao:2020nwy, Bedolla:2019zwg, Zhao:2020jvl,Yu:2022lak,Debastiani:DiquarkSize2017} and plot their Regge trajectories in Figs.~\ref{fig:libraryRegge_Swave}, \ref{fig:libraryRegge_Pwave}, and \ref{fig:libraryRegge_Dwave} identified by their $J^{PC}, \, L$, and $S$ numbers. 
Following the fits in Fig.~\insupp{5 in the main document}{\ref{Regge}}, we exclude the theoretical masses for the $n=1$ state from the fits. 
Each plot also includes our measured \PX trajectory (Fig.~\insupp{5 in the main document}{\ref{Regge}}).

The first three plots in Fig.~\ref{fig:libraryRegge_Swave} are for S-wave ($L=0$) configurations, and correspond to spin-0 diquarks ($0^{++}$), and spin-1 diquarks with anti-aligned spins ($0^{++}$) or aligned spins ($2^{++}$).
No $J=1$ result is shown as it results in negative charge conjugation for $L=0$ states, which is excluded. 
The plots in Fig.~\ref{fig:libraryRegge_Pwave} are for P-wave ($L=1$) and correspond to spin-1 diquarks with $J^{PC}=0^{-+}, 1^{-+}, 2^{-+}$. 
No spin-0 diquark results are shown as they produce a negative charge conjugation quantum number.
Figure~\ref{fig:libraryRegge_Dwave} is for the D-wave ($L=2$) prediction from Ref.~\cite{Bedolla:2019zwg}, and shows only the $0^{++}$ trajectory, because the squared masses of the $J=0, \, 1, \, 2$ states are indistinguishable on the scale of this plot. 
The $2^{++}$ state with $L=2$ arising from the spin-0 diquark model in Ref.~\cite{Bedolla:2019zwg} is also nearly identical to that from spin-1 diquarks and is therefore not plotted. 
Compared with $L=0$ states, theoretical predictions for the radial excitations of $L=1$ states are relatively scarce, and those for $L=2$ states are even more limited.
As also seen in Fig.~\insupp{5 in the main document}{\ref{Regge}}, trajectories obtained from the masses calculated in Ref.~\cite{Zhu:2020xni} for spin-1 diquarks under the S-wave assumption, in either a $0^{++}$ or $2^{++}$ state, are in good agreement with the \PX data.  
The other trajectories show a considerable range of slopes and intercepts, indicating the extensive variability in the current modeling of tetraquarks, even within the diquark-antidiquark framework.

%% file: BPH-24-003-public-authorlist.tex
\cmsinstitute{Yerevan Physics Institute, Yerevan, Armenia}
{\tolerance=6000
A.~Hayrapetyan, V.~Makarenko\cmsorcid{0000-0002-8406-8605}, A.~Tumasyan\cmsAuthorMark{1}\cmsorcid{0009-0000-0684-6742}
\par}
\cmsinstitute{Institut f\"{u}r Hochenergiephysik, Vienna, Austria}
{\tolerance=6000
W.~Adam\cmsorcid{0000-0001-9099-4341}, J.W.~Andrejkovic, L.~Benato\cmsorcid{0000-0001-5135-7489}, T.~Bergauer\cmsorcid{0000-0002-5786-0293}, M.~Dragicevic\cmsorcid{0000-0003-1967-6783}, C.~Giordano, P.S.~Hussain\cmsorcid{0000-0002-4825-5278}, M.~Jeitler\cmsAuthorMark{2}\cmsorcid{0000-0002-5141-9560}, N.~Krammer\cmsorcid{0000-0002-0548-0985}, A.~Li\cmsorcid{0000-0002-4547-116X}, D.~Liko\cmsorcid{0000-0002-3380-473X}, M.~Matthewman, I.~Mikulec\cmsorcid{0000-0003-0385-2746}, J.~Schieck\cmsAuthorMark{2}\cmsorcid{0000-0002-1058-8093}, R.~Sch\"{o}fbeck\cmsAuthorMark{2}\cmsorcid{0000-0002-2332-8784}, D.~Schwarz\cmsorcid{0000-0002-3821-7331}, M.~Shooshtari\cmsorcid{0009-0004-8882-4887}, M.~Sonawane\cmsorcid{0000-0003-0510-7010}, W.~Waltenberger\cmsorcid{0000-0002-6215-7228}, C.-E.~Wulz\cmsAuthorMark{2}\cmsorcid{0000-0001-9226-5812}
% ...
\par}
\vskip\cmsinstskip
\dag:~Deceased\\
$^{1}$Also at Yerevan State University, Yerevan, Armenia\\
$^{2}$Also at TU Wien, Vienna, Austria\\
% ...
